\documentclass[twocolumn,showpacs,prb,preprintnumbers,floatfix,amsmath,amssymb]{revtex4}
\usepackage{dcolumn}% Align table columns on decimal point
\usepackage{bm}% bold math
\usepackage{ulem}
\usepackage{graphicx}
\usepackage{eepic}
\begin{document}

\preprint{APS/123-QED}

\title{Real-Space Variational  Gutzwiller Wave functions for the Anderson-Hubbard Model}% Force line breaks with \\
%\email{avid@physics.queensu.ca}
\author{A. Farhoodfar$^{1}$}
\email{avid@physics.queensu.ca}
\author{X. Chen$^1$}
\author{R. J. Gooding$^{1}$}
\author{W. A. Atkinson$^{2}$}
\affiliation{$^1$Department of Physics Queen's University, Kingston ON K7L 3N6 Canada$\\$}
\affiliation{$^2$Department of Physics Trent University, Peterborough ON K9J 7B8 Canada$\\$}

\date{\today}% It is always \today, today,

\begin{abstract}
Partially-projected Gutzwiller variational wavefunctions are used to describe
the ground state of disordered interacting systems of fermions.  We compare several different variational ground states with the exact ground state for disordered one-dimensional chains, with the goal of determining a minimal set of variational parameters required to accurately describe the spatially-inhomogeneous charge densities and spin correlations.
We find that, for weak and intermediate disorder, it is sufficient to include spatial variations of the charge densities in the product state alone, provided that screening of the disorder potential is accounted for.  For strong disorder, this prescription is insufficient and it is necessary to include spatially inhomogeneous variational parameters as well.
\end{abstract}


\maketitle

\section{Introduction}
The problem of how to treat theoretically systems of interacting electrons in disordered materials extends over many decades.
The most difficult problems to solve are ones in which the physical properties of interest depend on spatial correlations of individual realizations of the disorder potential.  In this case, the majority of disorder-averaged approximations fail and more sophisticated analytical or numerical techniques are required.  Anderson localization and Coulomb gap physics are two examples of such problems\cite{Altshuler1985}.   The problems are particularly challenging in narrow band materials where conventional approximations for the Coulomb interaction are poor.\cite{imada98}

There has been particular interest in low-dimensional disordered systems in recent years.  For example, many transition metal oxides, especially the quasi-two-dimensional high temperature superconductors,\cite{kastner98,gooding97,lai98,Alloul2007,Sugimoto2006,Fujita2005,hashimoto2008,Ando03} are intrinsically disordered by chemical doping, and have electronic properties that appear to be strongly susceptible to disorder.  More recently, experiments have demonstrated the formation of a two-dimensional electron gas at the interface of an otherwise insulating La$_2$CuO$_4$/Sr$_2$TiO$_4$ heterostructure,\cite{Ohtomo2004,Reyren2007,Ueno2008,Caviglia2008}  and there are indications that the electronic states at the interface are strongly influenced by inhomogeneities.\cite{Schneider2008}  A more fundamental question that has arisen in the past decade is whether interactions may drive a delocalization transition in thin metal films.\cite{Kravchenko2004}

Theoretically, we can investigate the relation between disorder and electron-electron interactions by studying the Anderson-Hubbard (AH) Hamiltonian.   The AH model is the Hubbard model\cite{hubbard63} for a lattice with site energies chosen
from a random distribution
\begin{eqnarray}
{\mathcal{H}}&=&\sum_{i,\sigma}V_i{\hat n}_{i,\sigma}-t\sum_{\langle i,j \rangle,\sigma}{ c}^\dagger_{i,\sigma}{ c}_{j,\sigma}
%\\ \nonumber
%&&
+ U\sum_i{\hat n}_{i,\uparrow}{\hat n}_{i,\downarrow}
\label{eq:Ham_AndHub}
\end{eqnarray}
where $i,j=1,\dots, N$ denote the sites of the lattice, $\langle i,j \rangle$ implies
that $i$ and $j$ are nearest neighbours, ${ c}_{i,\sigma}$ (${\hat n}_{i,\sigma}$)
is the destruction (number) operator for an electron at site $i$ with spin $\sigma,$
and the hopping energy is denoted by $-t$. The on-site energy at site $i$ is given by
$V_i$, and $U$ is the electron repulsion of two electrons sharing
the same site.

 The only known exact solution for the AH Hamiltonian is in infinite-dimensions,\cite{georges96}  and we thus have to use approximate methods in finite dimensions.
The various approximate methods used previously include self-consistent Hartree-Fock theory,\cite{Milovanovic1989,Tusch1993,Heidarian2004,Fazileh2006,chen08} dynamical mean field theories,\cite{Ulmke1995,Laad2001,Nikolic2003,Byczuk2005,Balzer2005,Lombardo2006,Song2008,Andrade2008} and exact numerical calculations for small clusters.\cite{chen08,Kotlyar2001,Srinivasan2003,Chiesa2008}
Of particular relevance to the current work is a recent variational wavefunction approach  by Pezzoli {\it et al},\cite{pezzoli08} which we discuss below.

 Here, we consider a variational wavefunction approach,
based on a modification of the Gutzwiller wave function (GWF)\cite{gutzwiller64,gutzwiller65,yokoyama87a} to include spatial inhomogeneity. The GWF is the simplest variational wave function for Hubbard-type Hamiltonians and is given by
\begin{eqnarray}
|\Psi_{{GWF}}\rangle = \prod_{i}(1-(1-g)\hat{n}_{i\uparrow}\hat{n}_{i\downarrow})|\psi_\mathrm{ps}\rangle \label{psiG-P}
\label{psiG-GW_0}
\end{eqnarray}
where $|\psi_\mathrm{ps}\rangle$ is a reference product (i.e.\ Slater-determinant) state. This function was originally introduced
to study the correlations of the ground state of the Hubbard Hamiltonian. The
variational parameter $0\leq g \leq 1$ incorporates
the effect of the Hubbard repulsion between electrons of opposite spins on the same
site, and is obtained by minimizing the energy functional
\begin{equation}
E_\mathrm{GWF}=\frac{\langle\Psi_\mathrm{GWF}|{{\cal H}}|\Psi_\mathrm{GWF}\rangle}{\langle\Psi_\mathrm{GWF}|\Psi_\mathrm{GWF}\rangle}.
\end{equation}
We note that $g=1$ corresponds to an unprojected wavefunction, while $g=0$ corresponds to a fully projected wavefunction in which there are no doubly-occupied sites.  Physically, the $g=0$ projection captures an essential feature of the large-$U$ Hubbard model:  it generates local paramagnetic moments without breaking the spin-rotational invariance of the lattice.

The GWF is hard to treat analytically, even for the disorder-free Hubbard Hamiltonian where analytical
results have only been obtained in one\cite{metzner87,metzner88,gebhard87,gebhard88}
and infinite\cite{metzner89} dimensions.   The variational GWF is therefore primarily
a  numerical method.\cite{ceperley77,koch99,yokoyama87a,yokoyama87b}  Our goal
is to investigate the quality of a number of simple variational wavefunctions by comparison of the variational ground states with exact diagonalization calculations on small clusters.

The GWF has certain well-known limitations:\cite{yokoyama87a}  it fails in the disorder-free case, for example, to describe the Mott transition at half-filling in finite dimensions.  In general, the GWF can be improved by the addition of nonlocal Jastrow factors that describe the long-range interactions between charge-density excitations.\cite{yokoyama90,Capello2005}
 Recently, Pezzoli {\it et.~al.}\cite{pezzoli08} used variational methods to study the AH Hamiltonian in two dimensions.   Their particular focus was on states near the Mott transition, which necessitated the inclusion of the Jastrow factors.
 Furthermore,  they included spatial inhomogeneity in both the product state $|\psi_\mathrm{ps} \rangle$ and in their local variational factors (e.g. $g\rightarrow g_i$).   These extra degrees of freedom resulted in a large number of variational parameters (several hundred for a typical lattice) and required sophisticated minimization schemes for the energy functional.\cite{sorella05}  It is therefore worth asking under what circumstances we can safely reduce the number of variational parameters and still obtain reasonable results for the variational wavefunction.

In the current work, we examine trial wavefunctions with a small number of variational parameters.  Spatial inhomogeneity is incorporated in the product state wavefunctions $|\psi_\mathrm{ps}\rangle$, however, the variational parameters are taken to be spatially homogeneous.  Our main result is that for weak and intermediate disorder, this simple wavefunction gives a surprisingly good description of the ground state away from the Mott transition.  It is only for strong disorder that our simple ansatz for $|\Psi_\mathrm{GWF}\rangle$  breaks down.

The paper is organized as follows:  In Sec.~\ref{sec:method}, we describe the variational wavefunction in detail, and introduce a number of candidate states for $|\psi_\mathrm{ps}\rangle$.  In Sec.~\ref{sec:results}, we asess the quality of the different trial wavefunctions by comparing with both exact diagonalization and unrestricted Hartree-Fock (UHF) calculations.  The latter comparison is motivated by the fact that UHF has been, until recently, the standard numerical technique for studying disordered systems.  Furthermore, a recent work\cite{chen08} has suggested that UHF actually provides good quantitative results for some local physical quantities in disordered systems (namely that disorder improves the quality of the UHF approximation).  We discuss the strengths and limitations of the simple GWF in Sec.~\ref{sec:discussion}, and conclusions are given in Sec.~\ref{sec:conclusions}.

\section{Method: Gutzwiller variational Approach}
\label{sec:method}

In this section, we introduce a number of variational states $|\Psi_\mathrm{GWF}\rangle$ that will be used for calculations.  We want to keep the simple form of the projection operator in Eq.~(\ref{psiG-GW_0}), which means that spatial inhomogeneity is introduced entirely through the product state.  It is for this reason that we consider a variety of product states, with the goal of determining how large an effect the initial choice for $|\psi_\mathrm{ps}\rangle$ makes on the final $|\Psi_\mathrm{GWF}\rangle$.

\subsection{The Product States}

Assume we are seeking the $N_e$-electron variational ground state for an $N$-site lattice with a particular realization of the disorder potential.
We then take  $|\psi_\mathrm{ps}\rangle$ to be the $N_e$-electron ground state of an
$N$-site bilinear (effectively noninteracting) Hamiltonian ${\cal H}_\mathrm{bl}$ whose disorder potential is determined by the site energies of ${\cal H}$.  We examine three
possibilities: (i) ${\cal H}_\mathrm{bl}$ is the Anderson Hamiltonian, obtained by
setting $U=0$ in Eq.~(\ref{eq:Ham_AndHub}) and renormalizing the site energies with a variational screening parameter, (ii) ${\cal H}_\mathrm{bl}$ is the self-consistent  Hartree-Fock (HF) decomposition of $\mathcal{H}$, (iii) ${\cal H}_\mathrm{bl}$ is the HF decomposition of $\mathcal{H}$ with screened site energies.

Since ${\cal H}_\mathrm{bl}$ is bilinear, $|\psi_\mathrm{ps}\rangle$ can be represented by
\begin{equation}
\label{eq:genSCHF} |\psi_\mathrm{ps}\rangle~=~ \gamma_1^\dagger
\gamma_2^\dagger \gamma_3^\dagger\dots \gamma_{N_e}^\dagger|0\rangle
~=~\prod_{\alpha=1}^{N_e}~ \gamma_\alpha^\dagger|0\rangle,
\end{equation}
where $\alpha$ labels the lowest energy single-particle eigenstates of ${\cal H}_\mathrm{bl}$ ordered from lowest ($\alpha=1$) to highest ($\alpha=N_e$) energy, and $\gamma_\alpha^\dagger$'s are creation operators for these eigenstates. The $\gamma_\alpha^\dagger$'s can be related to the real-space creation operators, $c_I^\dagger$,
by:
\begin{equation}
\label{eq:productstates}
\gamma_\alpha^\dagger~=~\sum_{I=1}^{2N}~G_{I,\alpha}~c_I^\dagger~~,
\end{equation}
where $G_{I,\alpha}$ are elements of the unitary matrix  $\uuline G$ that diagonalizes ${\cal H}_\mathrm{bl}$,  $I$ labels the spin and site with $1~\leq ~I~\leq~2N$. Then we have
\begin{equation}
\label{eq:genSCHFexpanded1} |\psi_\mathrm{ps}\rangle~=~
\sum_{I_1,\dots ,I_{N_e}=1}^{2N} G_{I_1,1}\dots
G_{I_{N_e},N_e} c_{I_1}^\dagger
\dots c_{I_{N_e}}^\dagger |0\rangle,
\end{equation}
which can be rewritten, using the anti-commutation relations of the
creation operators, as
\begin{equation}
\label{eq:genSCHFexpanded2} |\psi_\mathrm{ps}\rangle=
\sum_{I_1<...<I_{N_e}}
                D[{\uuline G}(\uline I)]
c_{I_1}^\dagger  \dots
c_{I_{N_e}}^\dagger |0\rangle,
\end{equation}
where $D[{\uuline G}(\uline I)]$ is a Slater determinant
\begin{equation}
\label{eq:det11} D[{\uuline G}(\uline I)]~\equiv~
                \det \begin{pmatrix}
                  G_{I_1,1} &  \dots & G_{I_1,N_e} \\
                  G_{I_2,1} &  \dots & G_{I_2,N_e} \\
                  \dots &\dots &  \dots \\
                  \dots & \dots&  \dots \\
                  G_{I_{N_e},1} & \dots & G_{I_{N_e},N_e}~~ \\
                \end{pmatrix}.
\end{equation}
The vector $\uline I=(I_1,\dots ,I_{N_e})$ labels different configurations of the electrons. For a given $\uline I$, $D[{\uuline G}(\uline I)]$ gives the weight of that configuration of electrons in the product state. Using Eq. (\ref{psiG-GW_0}), we can write
\begin{eqnarray}
\label{eq:PGSCHFexpanded1}
&&|\Psi_\mathrm{GWF}\rangle=\\ \nonumber
\\\nonumber
&&\hspace{-0.6cm}\sum_{I_1<...<I_{N_e}}
\hspace{-0.4cm}D[\uuline G({\uline I})]
\prod_{i=1}^{N}(1-(1-g)\hat{n}_{i\uparrow}\hat{n}_{i\downarrow})
c_{I_1}^\dagger \dots
c_{I_{N_e}}^\dagger |0\rangle .
\end{eqnarray}
In general, $|\Psi_\mathrm{GWF}\rangle$ contains a large number of terms and must be calculated approximately, for example using variational Monte Carlo methods. In the current work, we restrict ourselves to small clusters where $|\Psi_\mathrm{GWF}\rangle$ can be evaluated exactly.

Our first choice of ${\cal H}_\mathrm{bl}$ is the non-interacting disordered Hamiltonian where the disorder potential is screened. The on-site energies are reduced,
\begin{equation}
{\underset{bare~disorder~potential}{\underbrace{V_i}}}\rightarrow{\underset{screened~disorder~potential}{\underbrace{V'_i=\frac{V_i}{\varepsilon}}}}\\\nonumber
\end{equation}
where the screening factor $\varepsilon$ is a variational parameter. Both variational parameters $\varepsilon$ and $g$ are determined so as to minimize the total GWF energy $E_\mathrm{GWF}(\varepsilon , g)$.
We refer to this variational wavefunction as the disordered fermi sea GWF (DFSGW).

Our remaining choices of ${\cal H}_\mathrm{bl}$ are based on the HF decomposition of ${{\cal H}}$. The four-fermion interaction term in Eq. (\ref{eq:Ham_AndHub}) can be written as
\begin{eqnarray}
\label{eq:HF_decomp}
U {\hat n}_{i\uparrow}{\hat n}_{i\downarrow}
&=& U({\overline n}_{i\uparrow}
+\delta {\hat n}_{i\uparrow})({\overline n}_{i\downarrow} +\delta {\hat n}_{i\downarrow})
\\ \nonumber
&\approx&U{\hat n}_{i\uparrow} {\overline n}_{i\downarrow} +
U{\hat n}_{i\downarrow} {\overline n}_{i\uparrow} -
U{\overline n}_{i\uparrow} {\overline n}_{i\downarrow},
\end{eqnarray}
where ${\overline n}_{i\sigma}\equiv\langle{\hat n}_{i\sigma}\rangle$, and $\delta \hat{n}_{i\sigma}=\hat{n}_{i\sigma}-{\overline n}_{i\sigma}$.\cite{NoteUHF} The HF Hamiltonian, using Eqs. (\ref{eq:Ham_AndHub}) and (\ref{eq:HF_decomp}), can be written as
\begin{equation}
{{\cal H}}_{HF}={\cal H}_\mathrm{bl}-U\sum_{i=1}^N\overline{n}_{i\uparrow}\overline{n}_{i\downarrow},
\end{equation}
where the second term on the right-hand side is just a constant, and
\begin{eqnarray}
{\cal H}_\mathrm{bl}=\sum_{i}\sum_{\sigma}(V_i+U\overline{n}_{i-\sigma})\hat{n}_{i\sigma}-t\sum_{\langle i,j \rangle\sigma} {c}^\dagger_{i\sigma}{c}_{j\sigma},
\label{Heff}
\end{eqnarray}
where $\overline{n}_{i\sigma}$ are determined self-consistently.

Our second choice of ${\cal H}_\mathrm{bl}$ is the paramagnetic HF (PMHF) Hamiltonian, i.e.\ Eq. (\ref{Heff}) with ${\overline n}_{i\uparrow}~=~{\overline n}_{i\downarrow}$.
Here $g$ is the only variational parameter, and we refer to the variational wavefunction for this choice as PMGW.

Our third choice of ${\cal H}_\mathrm{bl}$ is the PMHF Hamiltonian with a screened disorder potential. For this case, the on-site energies of Eq. (\ref{Heff}) are reduced to $V'_i=V_i/\varepsilon$, and again, both variational parameters $\varepsilon$ and $g$ are determined so as to minimize the total GWF energy.  These calculations are more computationally demanding than the previous two because $\overline{n}_{i\sigma}$ must be determined self-consistently for each $\varepsilon$. We refer to the variational wavefunction for this choice as PMGW$(g,\varepsilon)$.  The different ${\cal H}_\mathrm{bl}$ are summarized in Table~\ref{tab:one}.

%\begin{center}

\begin{table}[tb]
%\vspace{0.5cm}
\begin{tabular}{|l|c|c|}
  \hline
             & $U\bar{n}_{i\uparrow(\downarrow)}~~$   &$V'_i$                \\ \hline
  DFSGW& $~U\bar{n}_{i\uparrow(\downarrow)}=0~$  &$V'_i=V_i/\varepsilon$  \\ \hline
  PMGW& $U\bar{n}_{i\uparrow}=U\bar{n}_{i\downarrow}$    &$V'_i=V_i~~$            \\ \hline
  PMGW$(g,\varepsilon)$ & $U\bar{n}_{i\uparrow}=U\bar{n}_{i\downarrow}$    &$V'_i=V_i/\varepsilon$  \\
  \hline
  \end{tabular}
\caption{
Summary of the different bilinear Hamiltonians ${\cal H}_\mathrm{bl}$ used to
generate $|\psi_{ps}\rangle$.}
\label{tab:one}
\end{table}
As a simple illustration, in Appendix \ref{sec:Two-Site} we compare the different product states for a disordered two-site system. As in the disorder-free case, \cite{yokoyama90} the sample GWFs span the Hilbert space of a the two-site case and yield exact results.

\section{Results: Comparisons of Exact and Variational Quantities}
\label{sec:results}
In this section we show results for PMGW, PMGW$(g,\varepsilon)$, and DFSGW
variational states for small clusters where variational calculations may be compared to exact diagonalization calculations.  Motivated by the earlier work\cite{chen08} demonstrating that the UHF approximation (in which the self-consistently determined $\bar{n}_{i\sigma}$ may depend on $\sigma$) provides accurate charge densities in disordered systems, we also consider UHF as a benchmark for the various GWF calculations.

The quantities that we have calculated are as follows:
\begin{itemize}
\item{} We have calculated the absolute difference of the exact and the variational energies per site defined by
\begin{equation}
\label{eq:msqdiff-Egs}
\delta {E}_\mathrm{var}\equiv\frac{1}{N}|{E}^{\textrm{~var}}-{E}^{\textrm{~ex}}|.
\end{equation}
\item{}We have evaluated the magnitude of the overlap between
the exact and variational wave functions, given by
\begin{equation}
|\langle \Psi_{\textrm{ex}} | \Psi_{\textrm{{var}}} \rangle|.
\end{equation}
\item{} We have calculated the local charge densities according to
\begin{equation}
\label{eq:localCDdefn}
{\overline n}_i \equiv \langle \Psi | \sum_\sigma {\hat n}_{i,\sigma} | \Psi \rangle,
\end{equation}
where $| \Psi \rangle$ represents the exact or variational wave function.
\item{} We have calculated the local spin correlations for near-neighbour sites in one dimension,
\begin{equation}
\label{eq:SdotSdefn}
\langle{\bf S}_i\cdot
 {\bf S}_{i+1}\rangle\equiv\langle \Psi | { {\bf S}}_i\cdot
{ {\bf S}}_{i+1}  | \Psi \rangle.
\end{equation}
\item{}We have calculated the average absolute difference of charge densities given by
\begin{equation}
\label{eq:msqdiff-rhos}
\langle \delta {\overline {{n}}}\rangle\equiv\frac{1}{N}\sum_{i=1}^N~|{\overline n}_i^{\textrm{~var}}-{\overline n}_i^{\textrm{~ex}}|.
\end{equation}
\item{} We have calculated the average absolute difference of local spin correlations defined by
\begin{equation}
\label{eq:msqdiff-sisj}
\langle \delta {\overline{ {\bf S}\cdot {\bf S}'}}\rangle\equiv\frac{1}{N}\sum_{i=1}^N~|{\langle{\bf S}_i\cdot
 {\bf S}_{i+1}\rangle}^{\textrm{~var}}-{\langle{\bf S}_i\cdot {\bf S}_{i+1}\rangle}^{\textrm{~ex}}|.
\end{equation}
\end{itemize}

\subsection{Six-Site Cluster}
\label{sec:six}
\begin{figure}[tb]
\begin{center}
\includegraphics[clip=true,width=\columnwidth]{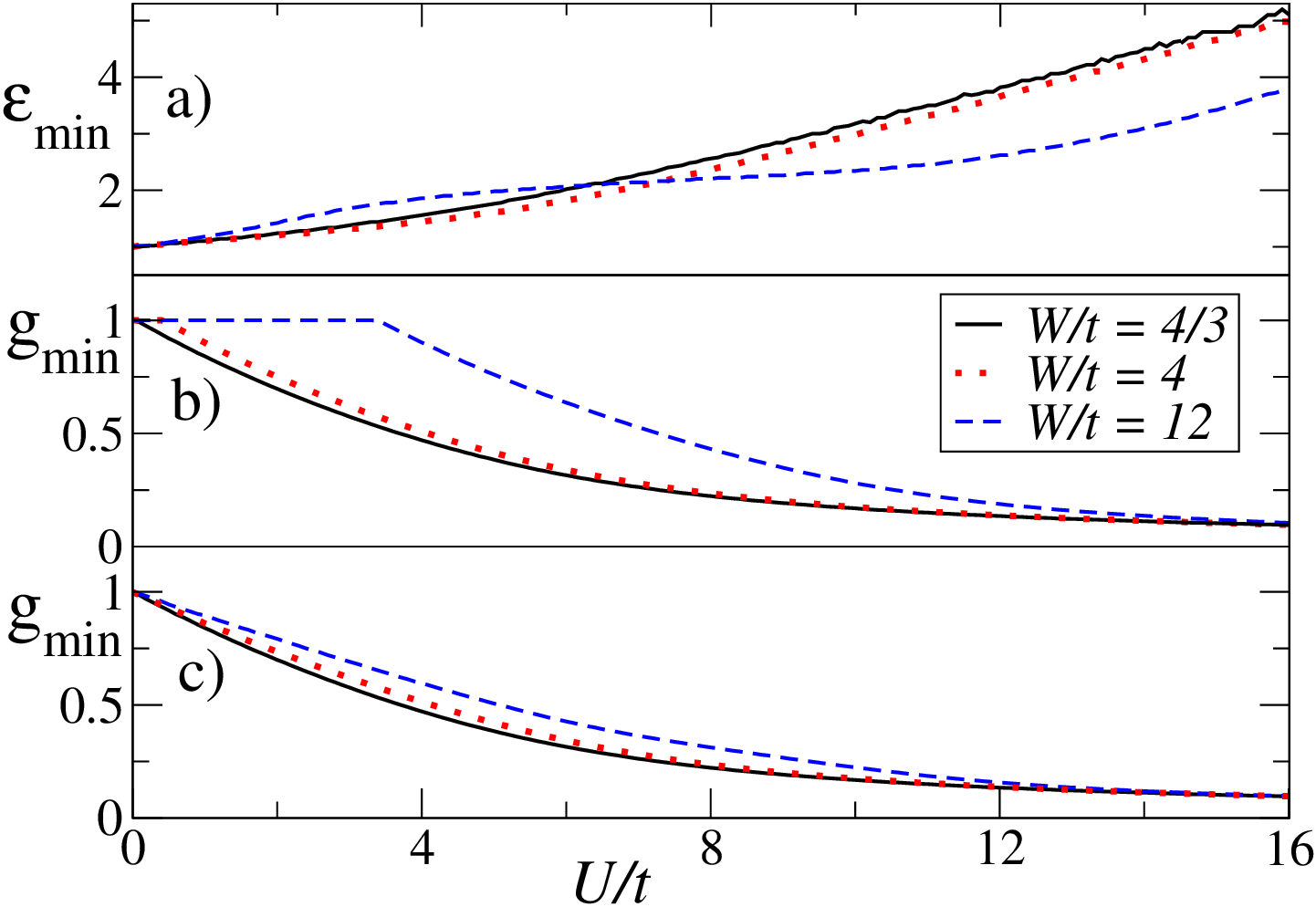}
\caption{\label{fig:screening_g_min}(Colour online)
Variational parameters for the DFSGW and PMGW approximations for the
six-site cluster for weak ($W/t = 4/3$), intermediate ($W/t = 4$), and strong disorder ($W/t = 12$).  The DFSGW
(a) screening factor $\varepsilon_{min}$,
and (b) projection $g_\mathrm{min}$ are shown as functions of $U$.  The
PMGW projection $g_\mathrm{min}$ is shown in (c).}
\end{center}
\end{figure}

\begin{figure}[tbp]
\begin{center}
\includegraphics[clip=true,width=\columnwidth]{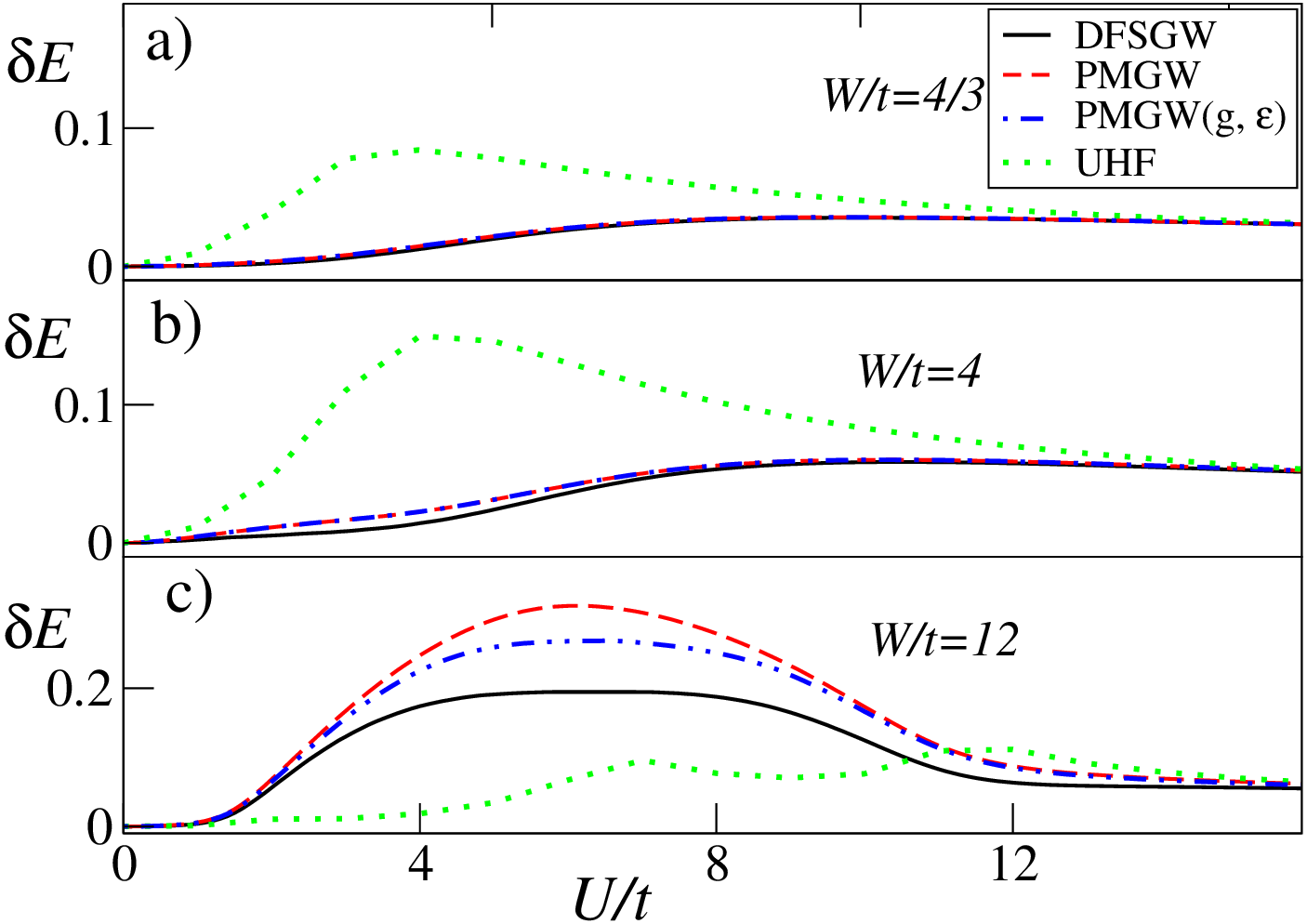}
\caption{\label{fig:6siteDIFnrgs}(Colour online)Difference between variational and exact energies
for (a) weak, (b) intermediate, and (c) strong disorder.
Results are shown for DFSGW, PMGW$(g,\varepsilon)$, PMGW, and UHF. The UHF results are shown for step size $\Delta U = t$. Note that for weak and intermediate disorder, the PMGW and
PMGW$(g,\varepsilon)$ curves coincide.}
\vspace{0.3cm}
\end{center}
\end{figure}

\begin{figure}[tb]
\begin{center}
\includegraphics[clip=true,width=\columnwidth]{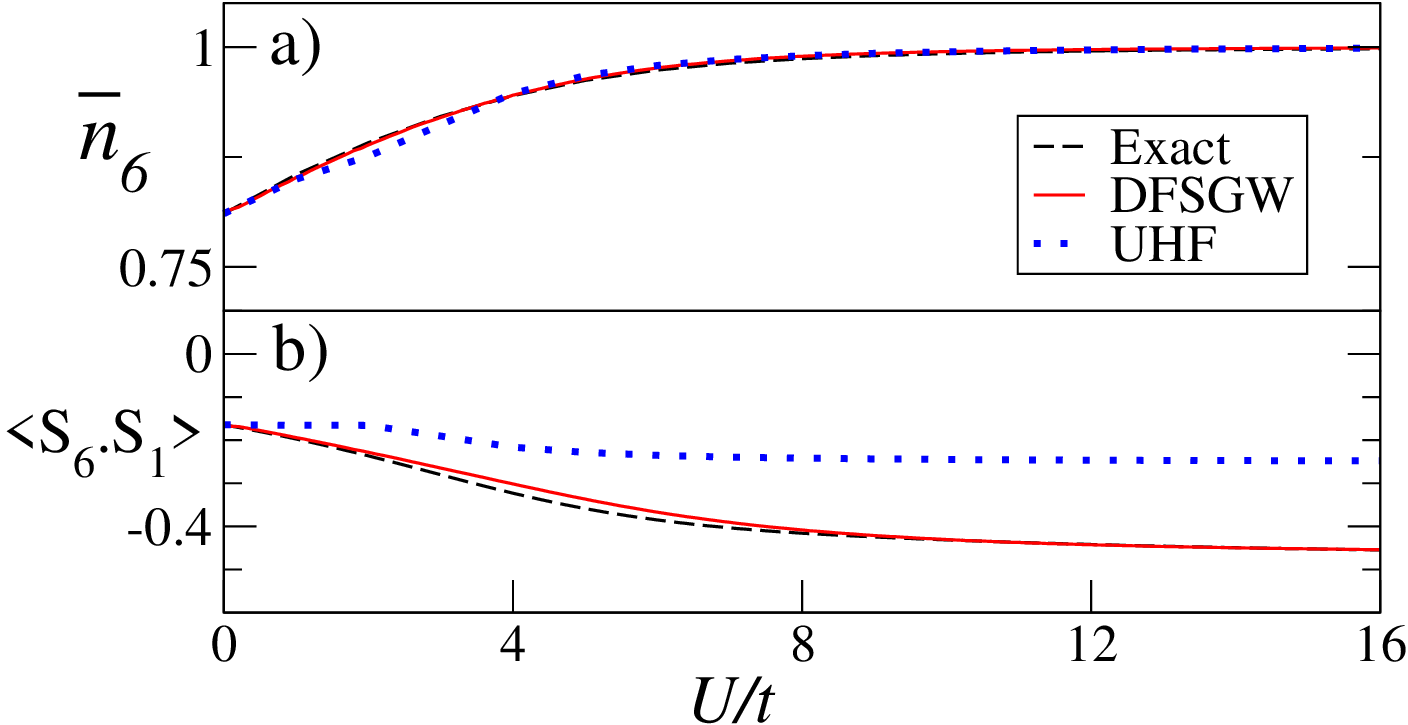}
\caption{\label{fig:comparisons_correlations_DFSGW_UHF_W4over3}(Colour online)
Comparison of (a) the charge density for the sixth site, $\bar{n}_6$,
and (b) $\langle{\bf S}_6\cdot {\bf S}_{1}\rangle$  for exact results (black dashed lines), DFSGW (solid red lines), and UHF(dotted blue lines). The UHF results are shown for step size $\Delta U = t$. Results are for $W/t=4/3$.}
\end{center}
\end{figure}

\begin{figure}[tbp]
%\vspace{0.4cm}
\begin{center}
\includegraphics[clip=true,width=\columnwidth]{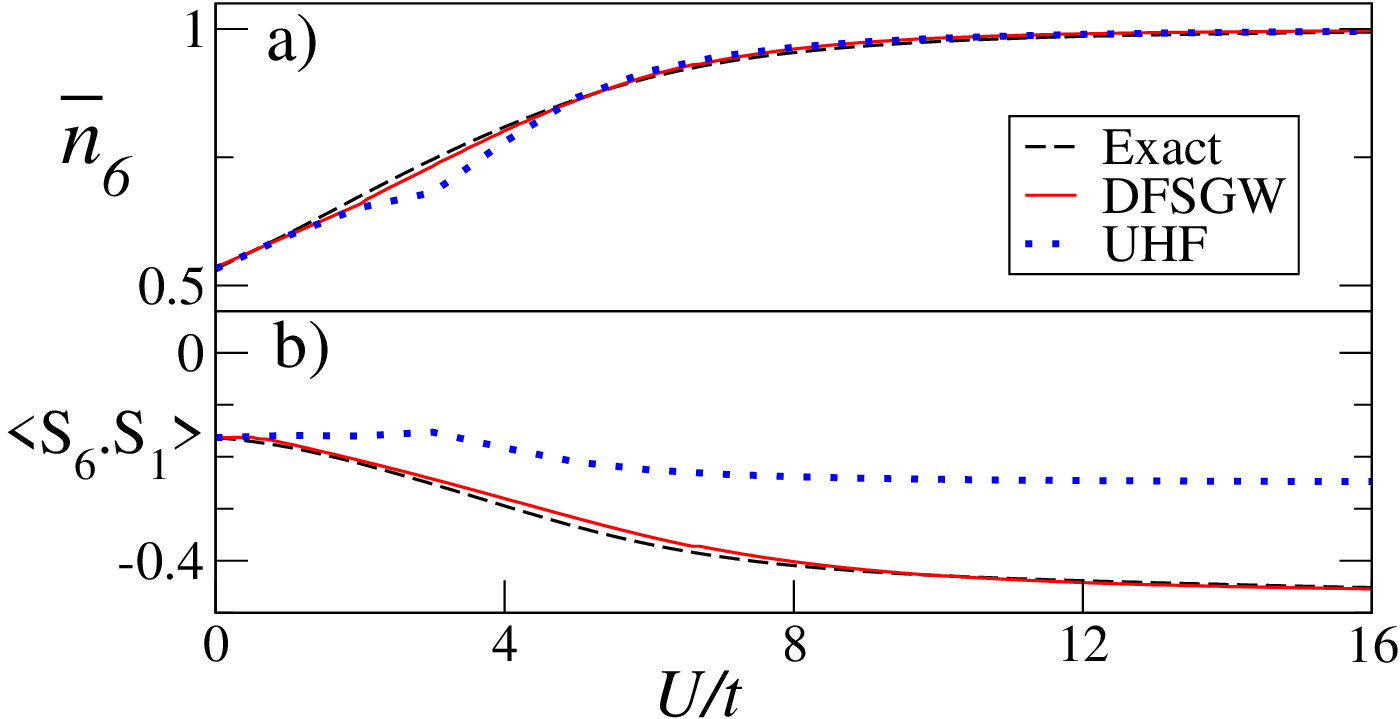}
\caption{(Colour online)As in Fig.~\protect\ref{fig:comparisons_correlations_DFSGW_UHF_W4over3}, but for $W/t=4$.}
\label{fig:comparisons_correlations_DFSGW_UHF_W4}
\end{center}
\end{figure}

\begin{figure}[tbp]
\begin{center}
%\vspace{0.4cm}
\includegraphics[clip=true,width=\columnwidth]{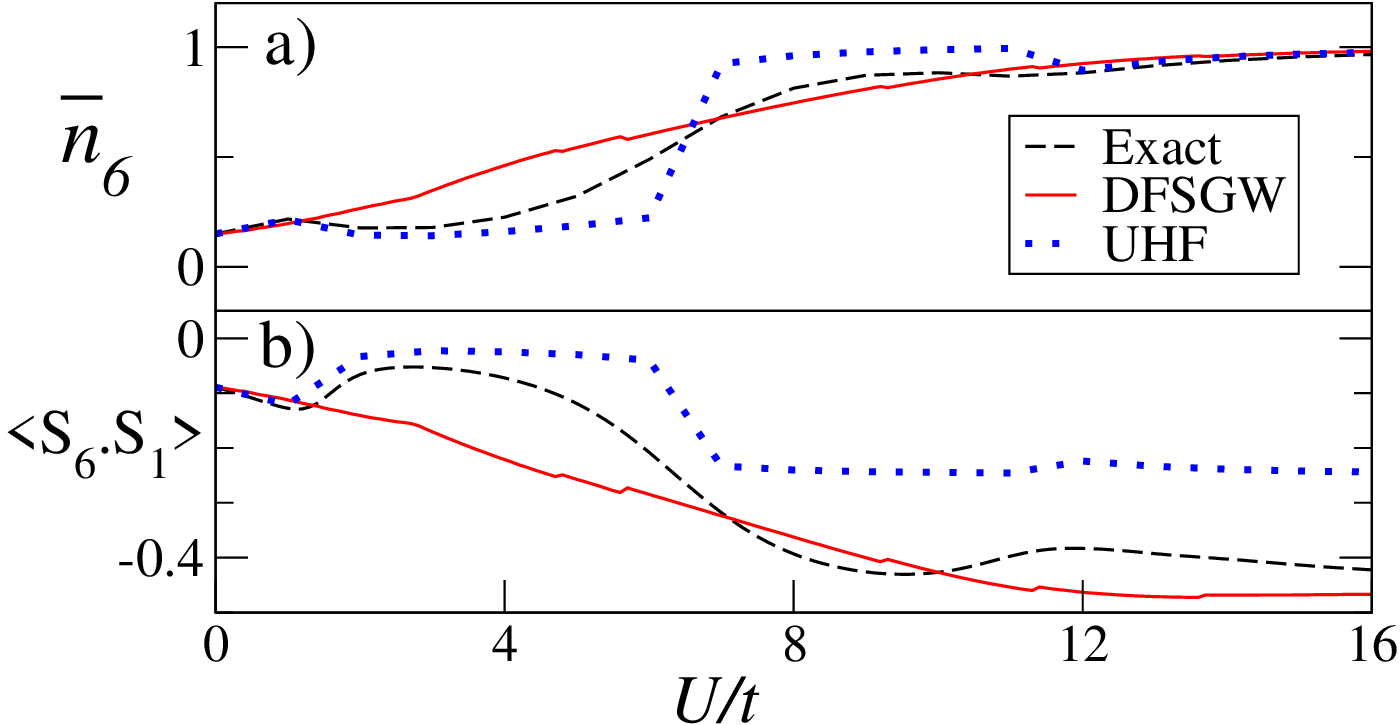}
\caption{\label{fig:comparisons_correlations_DFSGW_UHF_W12}(Colour online)
As in Fig.~\protect\ref{fig:comparisons_correlations_DFSGW_UHF_W4over3}, but for $W/t=12$.}
\end{center}
\end{figure}

\begin{figure}[tbp]
\begin{center}
\includegraphics[clip=true,width=\columnwidth]{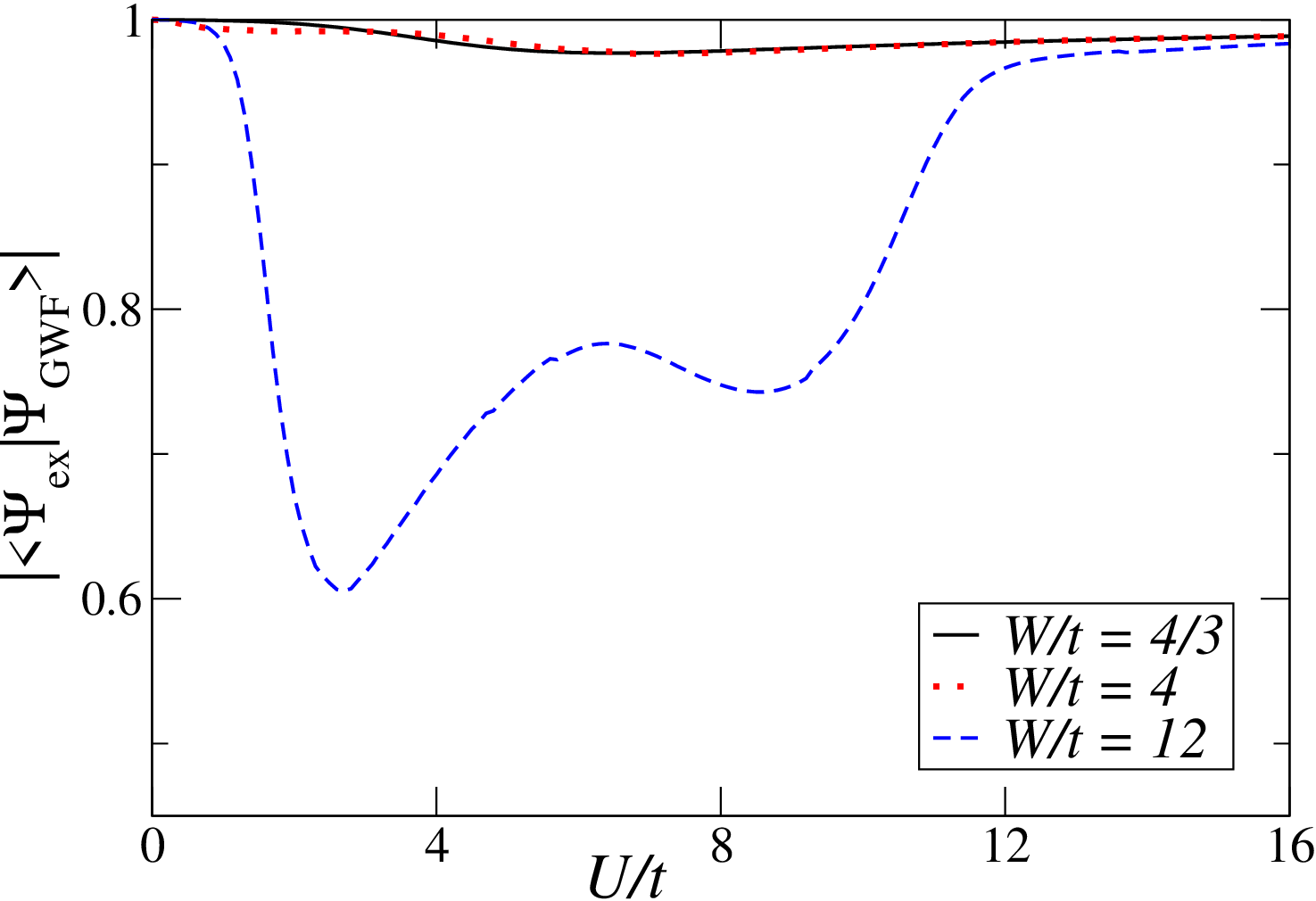}
\caption{\label{fig:6overlapDFSGW}(Colour online)
Magnitude of the overlap of the exact and DFSGW wave functions for the six-site disordered cluster.}
\end{center}
\end{figure}

\begin{figure}[tbp]
\begin{center}
%\vspace{0.2cm}
\includegraphics[clip=true,width=\columnwidth]{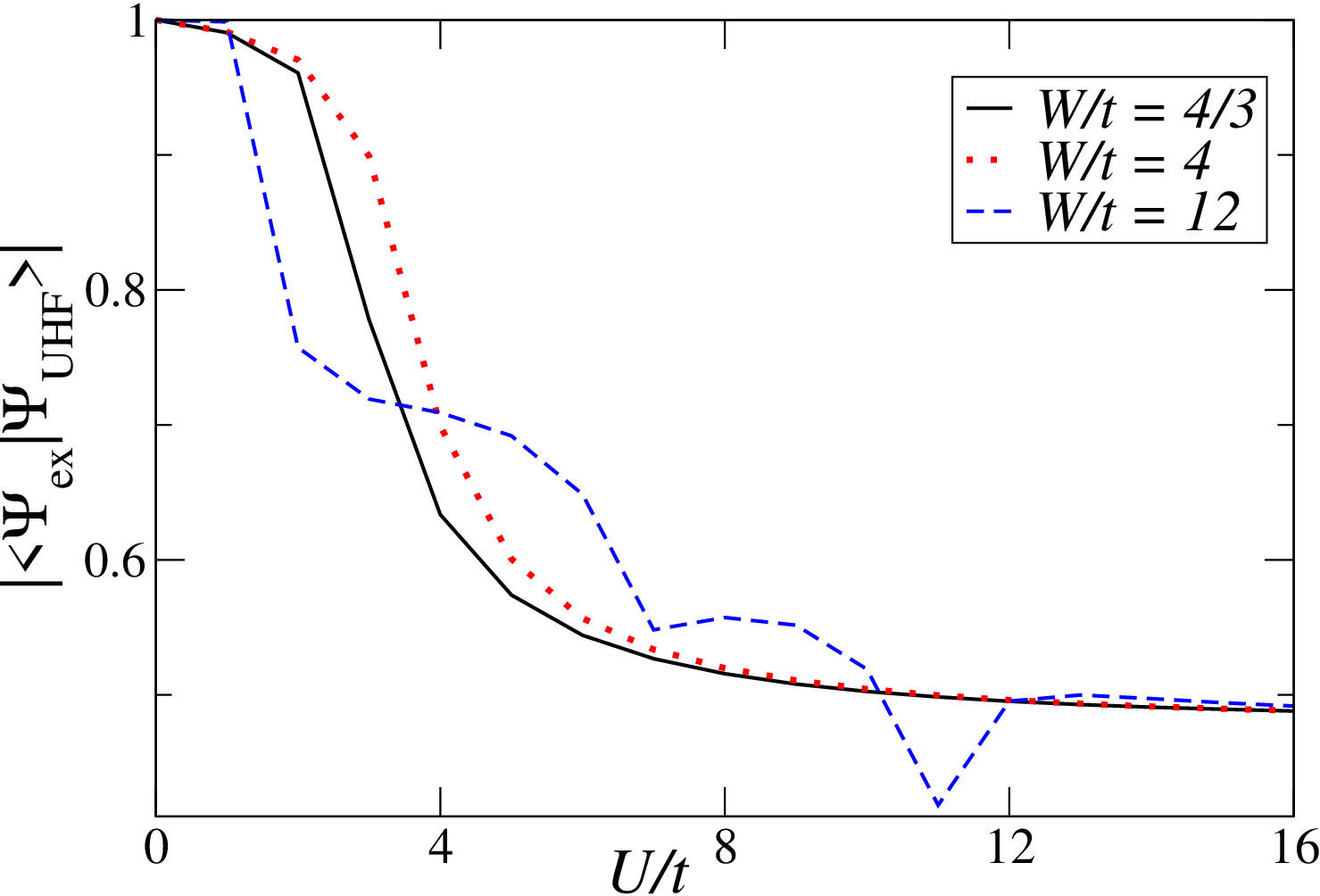}
\caption{\label{fig:6overlapUHF}(Colour online)
Magnitude of the overlap of the exact and UHF wave functions for the six-site disordered cluster. Data are shown for step size $\Delta U = t$}
\end{center}
\end{figure}

We have examined several complexions of disorder for the $6\times 1$ cluster with periodic boundary conditions at half-filling ($N_e=6$), and focus on one representative
configuration. The disorder potential for the configuration is
\begin{equation}
\label{eq:6sitedisorder}
V_i~=~0,~-0.18W,~+0.5W,~+0.12W,~-0.5W,~+0.3W,\nonumber
\end{equation}
where $W$ is the strength of disorder. The bandwidth $D=4t$ for the noninteracting ordered cluster sets the scale for the disorder potential, and we have examined weak ($W/t = 4/3$), intermediate ($W/t = 4$) and strong ($W/t = 12$) disorder.
Our results are shown in Figs. \ref{fig:screening_g_min} to \ref{fig:6overlapUHF}.

We show first, in Fig.~\ref{fig:screening_g_min}, the variational parameters $g_{min}$ and $\varepsilon_{min}$ that minimize the total energy for the DFSGW and PMGW approximations.   The PMGW$(g,\varepsilon)$ approximation is numerically equivalent to PMGW (i.e.\ $\varepsilon_{min}=1$) for weak and intermediate disorder, and is therefore not shown.

For the DFSGW approximation, $\varepsilon_{min}$ is an increasing function of $U$, indicating that interactions effectively screen the disorder potential, making the charge distribution more homogeneous than in the noninteracting case.   At weak and intermediate disorder, $\varepsilon_{min}$ is only weakly affected by the disorder potential;  however, for strong disorder there is a significant reduction in screening.   For all $W$,   $\varepsilon_{min} \propto U$ when $U>W$.  For weak disorder, $g_\mathrm{min}$ is quantitatively like that of the ordered case,\cite{yokoyama87a}
while for strong disorder the wavefunction is unprojected ($g_\mathrm{min}=1$) for $U<3.4t$.   In all cases, the projection is substantial when $U\gg W,D$.

For the PMGW approximation, the screening is implicit in the self-consistency of charge densities in the HF product state.  For weak and intermediate disorder, $g_\mathrm{min}$ is quantitatively similar to the DFSGW case; however, for strong disorder $g_\mathrm{min}<1$ for all nonzero $U$, in contrast to the DFSGW case.  The quantitative similarity between the different $g_\mathrm{min}$ curves in Fig.~\ref{fig:screening_g_min}(c), unlike Fig.~\ref{fig:screening_g_min}(b), suggests that the self-consistent HF solutions screen the disorder potential more completely than the variational parameter $\varepsilon_{min}$ in the DFSGW approximation.

We now move to a discussion of the quality of the variational solutions.  In Fig. \ref{fig:6siteDIFnrgs}, we show the differences between the exact and variational energies. For weak and intermediate disorder $\delta E_\mathrm{GWF}$ is smaller than $\delta E_\mathrm{UHF}$, suggesting that GWFs are better than UHF.
Both the DFSGW and PMGW variational states have similar  $\delta E$ (recall that PMGW$(g,\varepsilon)$ is numerically equivalent to PMGW here), and detailed comparisons  of the variational states (not shown) find little difference in their predictions for the physical observables defined at the beginning of Sec.{~III}.

For strong disorder the situation is a little different:  the value for $\delta E_\mathrm{GWF}$ is different for the three variational states,
with DFSGW having the smallest $\delta E$.
However, for most values of $U$, the UHF approximation
has a lower energy than any of the GWF approximations.

In Figs~\ref{fig:comparisons_correlations_DFSGW_UHF_W4over3} to \ref{fig:comparisons_correlations_DFSGW_UHF_W12} we compare charge densities and spin correlations for variational and exact calculations.
 For clarity,  GWF results are shown only for the DFSGW approximation.
 Furthermore, we show the charge density as a function of $U$ for only one representative site in the cluster, and we show the spin correlations for only a single representative pair of sites.

Figure \ref{fig:comparisons_correlations_DFSGW_UHF_W4over3} shows the results for weak disorder, $W/t=4/3$.
Here, $W\ll D$, and the bandwidth $D$ is therefore the relevant energy scale for the crossover between the weakly and strongly-interacting limits as a function of $U$ (similar to the disorder-free case\cite{yokoyama87a}).
The figure shows that  both the UHF and DFSGW do a good job of reproducing the local charge densities for all $U$, but that the DFSGW is significantly better at reproducing the spin-correlations.  This is particularly true at large $U$ where local moments have formed (i.e. all sites are singly occupied): within the UHF approximation, local moment formation coincides with the onset of classical static magnetic moments, such that $\langle {\bf S}_i\cdot {\bf S}_{i+1} \rangle\rightarrow -1/4$. In contrast, the DFSGW gives $\langle {\bf S}_i\cdot {\bf S}_{i+1} \rangle\rightarrow -0.45$ for large $U$, which is the clean limit result for large $U$\cite{kaplan82}.

Results for intermediate disorder are shown in Fig.~\ref{fig:comparisons_correlations_DFSGW_UHF_W4}, and the conclusions are the same as for weak disorder:  both the UHF and the DFSGW do a good job of reproducing the local charge density, but the UHF fails to reproduce the spin correlations.  A closer examination reveals that the UHF value for $\bar{n}_6$ deviates from the exact $\bar{n}_6$ near $U=1.3t$,
 where local moments form at site 6.

  By contrast, results for large disorder
in Fig.~\ref{fig:comparisons_correlations_DFSGW_UHF_W12} show that while the DFSGW captures general trends, it does a poor job of reproducing the details of the charge densities and spin correlations as a function of $U$.  Interestingly, the UHF does a remarkably good job of describing both the charge densities and spin correlations for $U\lesssim 6t$.   We discuss the large-disorder limit in more detail below, but remark here that the success of UHF can be traced back to the fact that, for $U\ll W$, most sites are empty or doubly occupied, with relatively few having moments.  The UHF approximation correctly predicts that spin correlations between isolated moments and their near-neighbors vanish when the near-neighbor sites are empty or doubly occupied.


Another measure of the quality of the various approximations is the wavefunction overlap, shown in Figs.~\ref{fig:6overlapDFSGW} and \ref{fig:6overlapUHF}.   The results of these figures are essentially consistent with the results presented above:  the DFSGW has a large overlap with the exact wave function ($|\langle \Psi_{ex}|\Psi_\mathrm{GWF}\rangle| > 0.977$)  for weak and intermediate disorder, while for strong disorder the overlap is poor except in the small- and large-$U$ limits.  The UHF overlap is generally poor, except at small $U$.

The reason for the difference in the quality of the GWF approximation at intermediate and large disorder is most easily understood by first considering the atomic limit ($t=0$).  When $U=0$, the lowest-energy $N_e/2$ sites are doubly occupied while the remaining sites are empty.  As $U$ is increased, this arrangement persists until $U$ is larger than the energy difference between the highest-energy doubly-occupied site and the lowest-energy empty site, at which point one electron is transferred from the doubly-occupied site to the empty site.  The process continues as $U$ increases further, with electrons being promoted from doubly-occupied sites whenever the cost of double occupancy is greater than the cost of promoting an electron to the next available empty site.  The process terminates when $U>W$, at which point all lattice sites are singly occupied.  The formation of local moments (singly-occupied sites) therefore happens inhomogeneously in the atomic limit.

A nonzero $t$ delocalizes electrons by an amount proportional to $t/W$, and thus makes the charge distribution more homogeneous.
Given that the Gutzwiller projection (which is responsible for generating local moments in $|\Psi_\mathrm{GWF}\rangle$) is chosen to be spatially uniform, it is unsurprising that the DFSGW approximation should work well for weak disorder ($W\ll D$)
and fail for strong disorder ($W\gg D$) where the physics approaches the atomic limit.    The surprising result, evident in
Fig.~\ref{fig:comparisons_correlations_DFSGW_UHF_W4},
is that the moment formation is sufficiently homogeneous at intermediate disorder ($W=D$) to be well-represented by our simple variational wavefunction.

In part, the success of the simple GWF approximation at intermediate disorder can be attributed to the screening of the disorder potential by interactions (c.f.~ Fig.~\ref{fig:screening_g_min}). For small $U$, correlation effects are minor and the wavefunction is well represented by the original product state $|\psi_\mathrm{ps}\rangle$, while for $U> D$ (where correlation effects are important), $\varepsilon_{min}$ produces a significant renormalization of the impurity potential.

\subsection{Ten-Site Cluster}
\label{sec:ten}
We now extend the work of the previous section to consider 10-site clusters with periodic boundary conditions.   In this section, we consider results which are averaged over 10 randomly-generated complexions of the disorder potential.  The site potentials are chosen to lie in the interval $(-W/2,W/2)$ and we consider weak ($W/t=4/3$), intermediate ($W/t=4$), and strong disorder ($W/t=12$) cases as before.
We show data for $N_e=10$ electrons (half-filling) and $N_e=6$ electrons (near to quarter-filling).   Given the large Hilbert space (63~500 states for $N_e=10$), we use the Lanczos algorithm to find the exact ground states for comparison to the GWF results.

Results are shown for the DFSGW and PMGW approximations, as well as for UHF.  Results have not been shown for the PMGW$(g,\varepsilon)$ approximation because it was found in Sec.~\ref{sec:six} to be identical to the PMGW approximation for weak and intermediate disorder, and because it is not significantly better than PMGW at large disorder.

%Error bars are shown for DFSGW results and are similar for other approximations. Error bars give the root mean square (rms) variation of the value over the 10 impurity configurations, and are shown for every $4th$ data point.
Throughout this section, we show error bars for DFSGW results. These error bars give the root mean square (rms) variation of $\varepsilon_{min}$, $g_\mathrm{min}$, $\langle\delta E\rangle$, $\langle\delta \overline{n}\rangle$, and $\langle \delta {\overline{{\bf S}\cdot{\bf S'}}}\rangle$ over $10$ impurity configurations and (for $\langle\delta \overline{n}\rangle$ and $\langle \delta {\overline{{\bf S}\cdot{\bf S'}}}\rangle$) $10$ sites. The bars are shown for every $4th$ data point and are of similar size in other GWF approximations. We emphasize that these error bars do not indicate the accuracy of the approximation
(the curves themselves indicate this), but describe the site-to-site or
sample-to-sample variation of the accuracy of the
approximation. Thus, small error
  bars indicate that the approximation consistently
  over/underestimates a quantity by the same amount, while large error
  bars indicate that the quality of the approximation varies
  significantly from site to site.

\subsubsection{$N_e=10$ electrons (half-filling)}

\begin{figure}[tbp]
\begin{center}
\includegraphics[clip=true,width=\columnwidth]{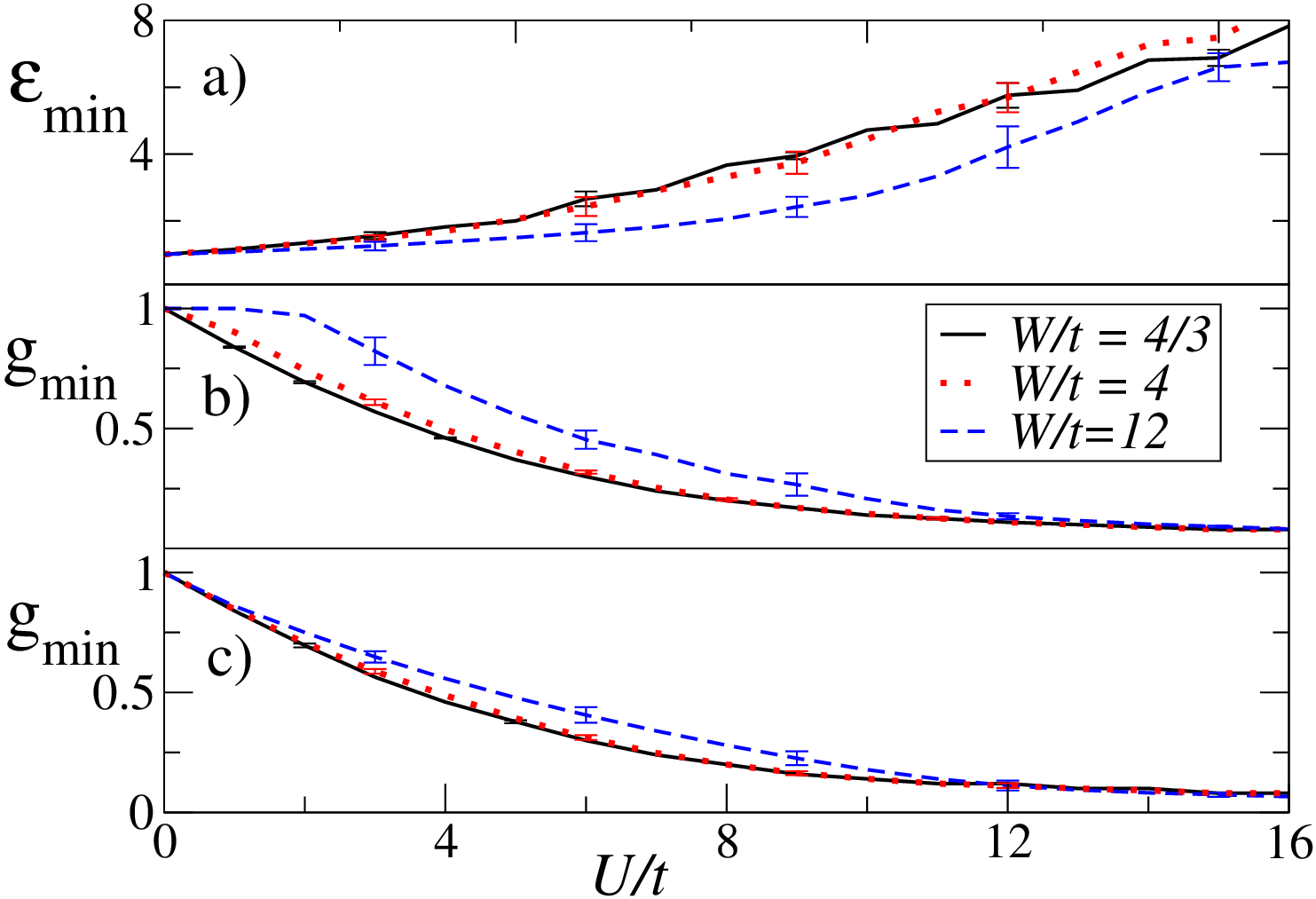}
\caption{\label{fig:g_vs_U_L10Ne10}(Colour online)
Variational parameters,  (a) and (b), for the DFSGW and  (c) PMGW approximations.  Results are shown for
ten-site disordered clusters averaged over ten complexions of disorder with $N_e=10$. Data are shown for step size $\Delta U = t$. Error bars give the rms variation of $\varepsilon_{min}$ and $g_{min}$ over impurity configurations.}
\end{center}
\end{figure}

We begin by showing results for the disorder-averaged variational parameters
as a function of $U$ at half-filling (Fig.~\ref{fig:g_vs_U_L10Ne10}).  The DFSGW curves are very similar to those shown in Fig.~\ref{fig:screening_g_min} for six sites.  As before, differences between the PMGW and DFSGW curves are most pronounced for large disorder and suggest that the PMHF approximation leads to a more complete screening of the disorder potential than does DFSGW.

\begin{figure}[tbp]
\begin{center}
%\vspace{0.3cm}
\includegraphics[clip=true,width=\columnwidth]{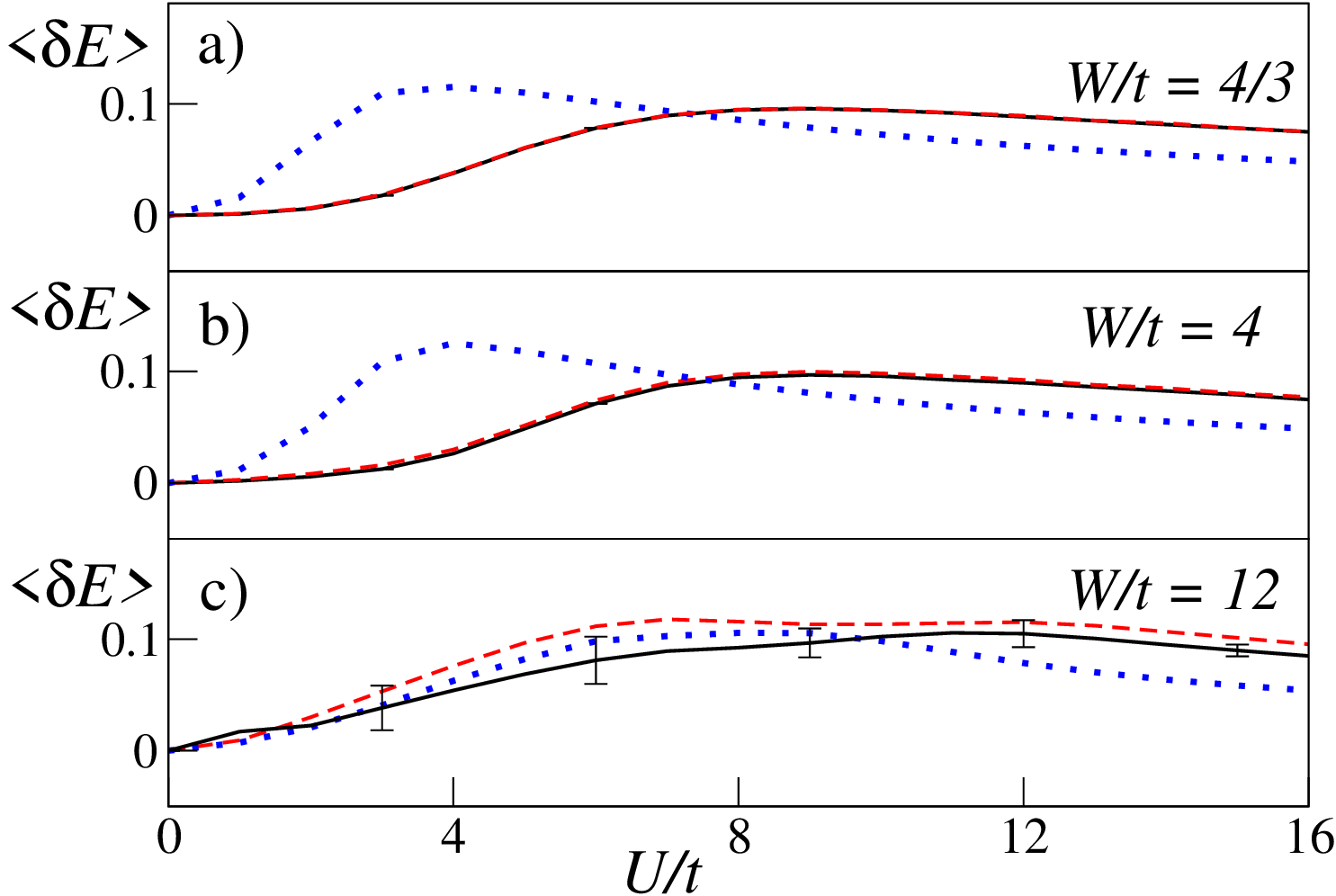}
\caption{\label{fig:10siteDIFnrgs10rand}(Colour online)
The disorder-averaged energy difference per site between the variational and exact energies for (a) weak, (b) intermediate, and (c) strong disorder at half-filling ($N_e=10$).  The curves are DFSGW (solid black line), PMGW (red dashed line), and UHF (dotted blue line). Data are shown for step size $\Delta U = t$. Error bars give the rms variation of $\langle \delta E\rangle$ over impurity configurations, and are too small to see in $(a)$ and $(b)$.}
\end{center}
\end{figure}

The variational energies are shown in
Fig.~\ref{fig:10siteDIFnrgs10rand}.
 For weak and intermediate disorder, the two GWF approximations give nearly identical values for $\langle\delta E\rangle_\mathrm{GWF}$.  The GWF energies are lower than the UHF energies for small $U$, but larger than the UHF energies at large $U$.  This is consistent with the results for larger clusters in the disorder-free case (c.f. Fig.~4 of Ref. 48), and is an indication that the simple GWF is unable to correctly reproduce the Mott transition at large $U$. For strong disorder, the Mott transition occurs at $U \approx W$ and this sets the energy scale at which the UHF becomes superior to the GWF.

\begin{figure}[tbp]
\begin{center}
\includegraphics[clip=true,width=\columnwidth]{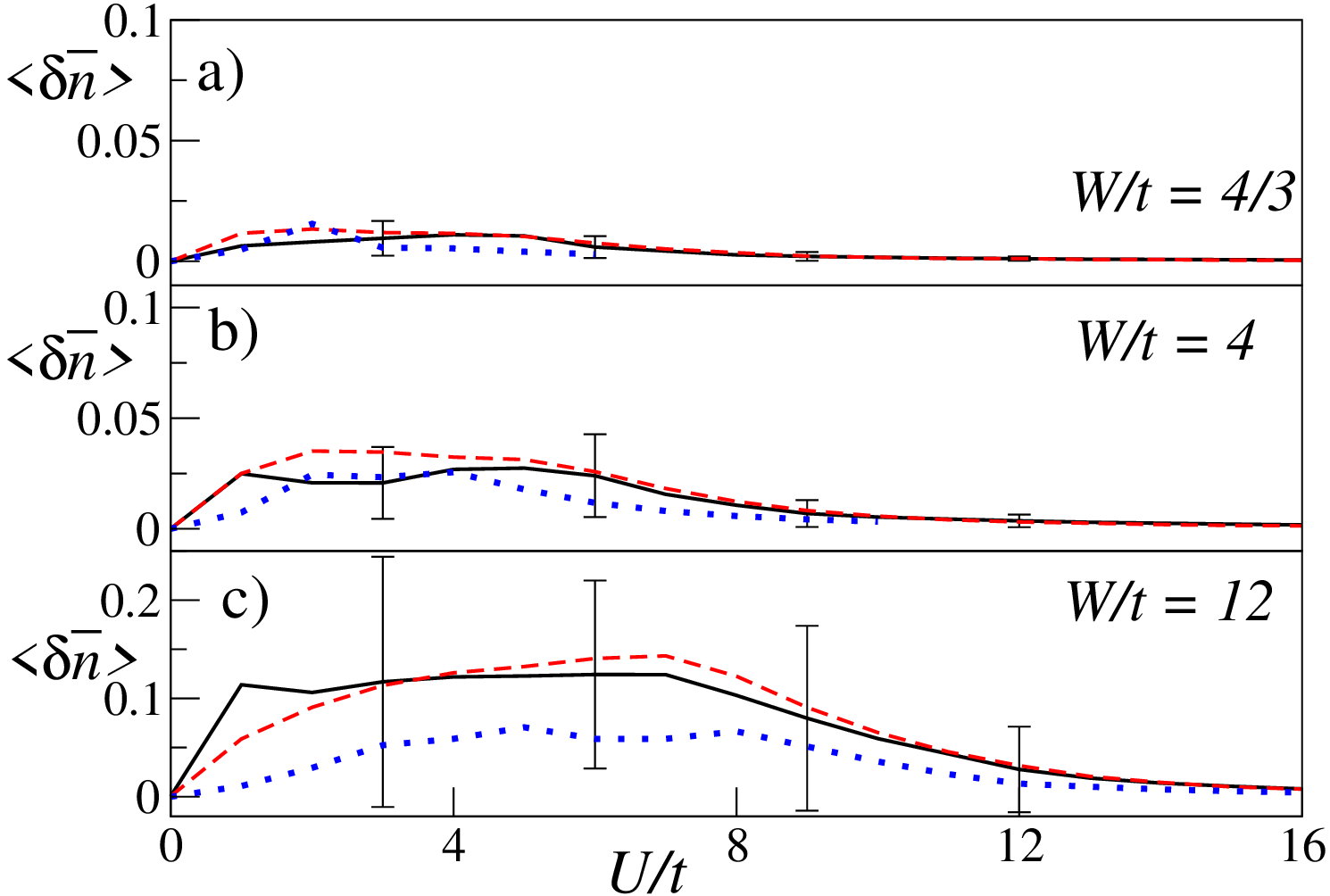}
\caption{\label{fig:10Ni10rand}(Colour online)
A comparison of the mean absolute difference between the
exact and variational local charge densities for the
ten-site disordered cluster averaged over ten complexions of disorder,
with $N_e=10$. Curves are for DFSGW (solid black line), PMGW (red dashed line), and UHF (dotted blue line). Data are shown for step size $\Delta U = t$. Error bars give the rms variation of $\langle \delta \overline{n}\rangle$ over sites and impurity configurations.}
\end{center}
\end{figure}

The error in the charge density $\langle \delta {\overline n}\rangle$, defined in Eq.~ (\ref{eq:msqdiff-rhos}), is shown in Fig.~\ref{fig:10Ni10rand}. At half-filling, the results are largely consistent with those found for six sites.  Both the GWF and UHF approximations are good for weak and intermediate disorder, with the DFSGW approximation producing an average error of less than 2.5\%; however,
all approximations work less well for strong disorder.

\begin{figure}[tbp]
\begin{center}
%\vspace{0.2cm}
\includegraphics[clip=true,width=\columnwidth]{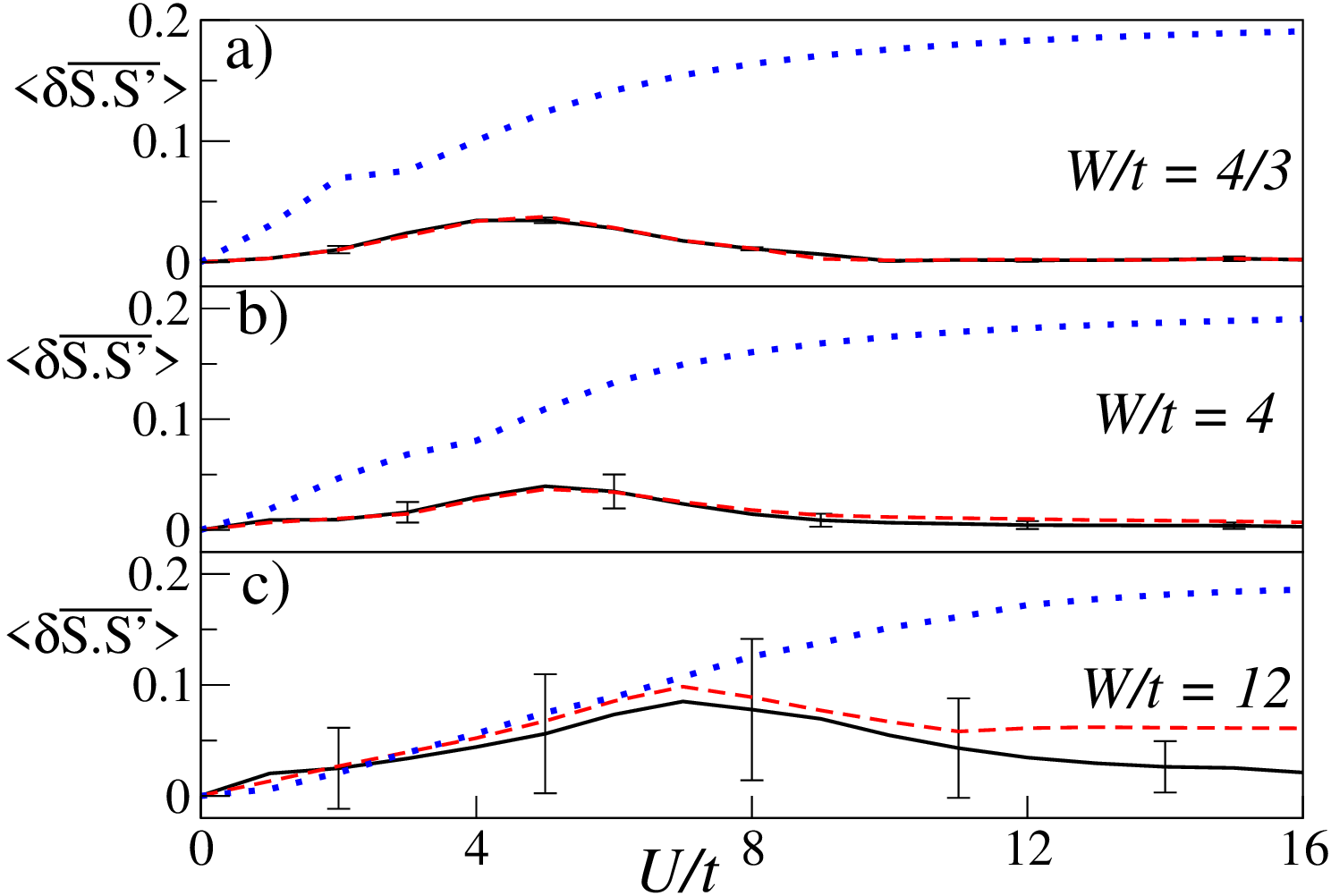}
\caption{\label{fig:10sisj10rand}(Colour online)
A comparison of the mean absolute difference between the
exact and variational expectation values of the near-neighbour spin correlations for the
ten-site disordered cluster averaged over ten complexions of disorder,
with $N_e=10$. Curves are for DFSGW (solid black line), PMGW (red dashed line), and UHF (dotted blue line). Data are shown for step size $\Delta U = t$. Error bars give the rms variation of $\langle \delta \overline{S\cdot S'}\rangle$ over sites and impurity configurations.}
\end{center}
\end{figure}

The spin correlations, defined in Eq.~(\ref{eq:msqdiff-sisj}), are shown in Fig.~\ref{fig:10sisj10rand} and are also similar to the six-site case.  The UHF approximation does a poor job because it generates static local moments, while the GWF approximations work remarkably well for weak and intermediate disorder, having errors of order 1-2\%.  For strong disorder, the results are comparable for all approximations, when $U~\lesssim ~7t$. The GWF approximations are much better than UHF for $U~>~7$.

\subsubsection{$N_e=6$ electrons}
\begin{figure}[tbp]
\begin{center}
\includegraphics[clip=true,width=\columnwidth]{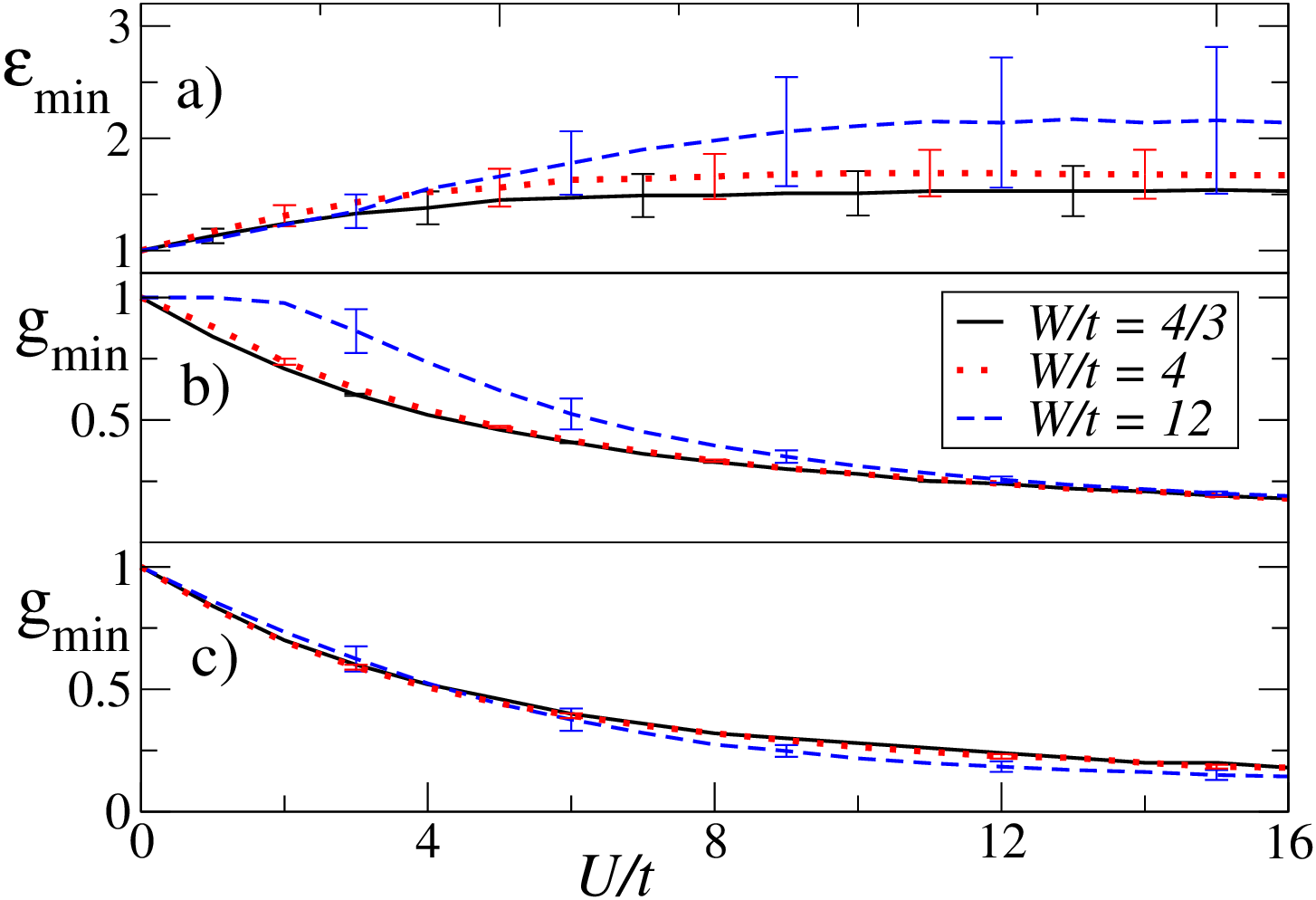}
\caption{\label{fig:g_vs_U_L10Ne6}(Colour online)
As in Fig.~\ref{fig:g_vs_U_L10Ne10}, but for $N_e=6$ electrons.}
\end{center}
\end{figure}
Away from half-filling, strong correlations play a lesser role than at half-filling, and there is no Mott transition.  The variational parameters for $N_e=6$ are shown as a function of $U$ in  Fig.~\ref{fig:g_vs_U_L10Ne6}.  The most noticeable difference with the half-filled case is that the variational parameters saturate at large $U$ here.  The energy scale at which saturation occurs appears to be $D$ for weak and intermediate disorder, and $W$ for strong disorder.   A consequence of saturation is that the impurity potential is only partially screened at large $U$.  Thus, while the half-filled ground state is the same for ordered and disordered models as $U\rightarrow \infty$, the disorder potential remains relevant at all $U$ for $N_e=6$.   This relevance is illustrated with a simple example:  for $W\gg t$ and $U\rightarrow \infty$,  the ground state has the $N_e$ sites with the lowest potentials $V_i$ singly-occupied and the remaining $N-N_e$ empty.
\begin{figure}[tbp]
\begin{center}
%\vspace{0.2cm}
\includegraphics[clip=true,width=\columnwidth]{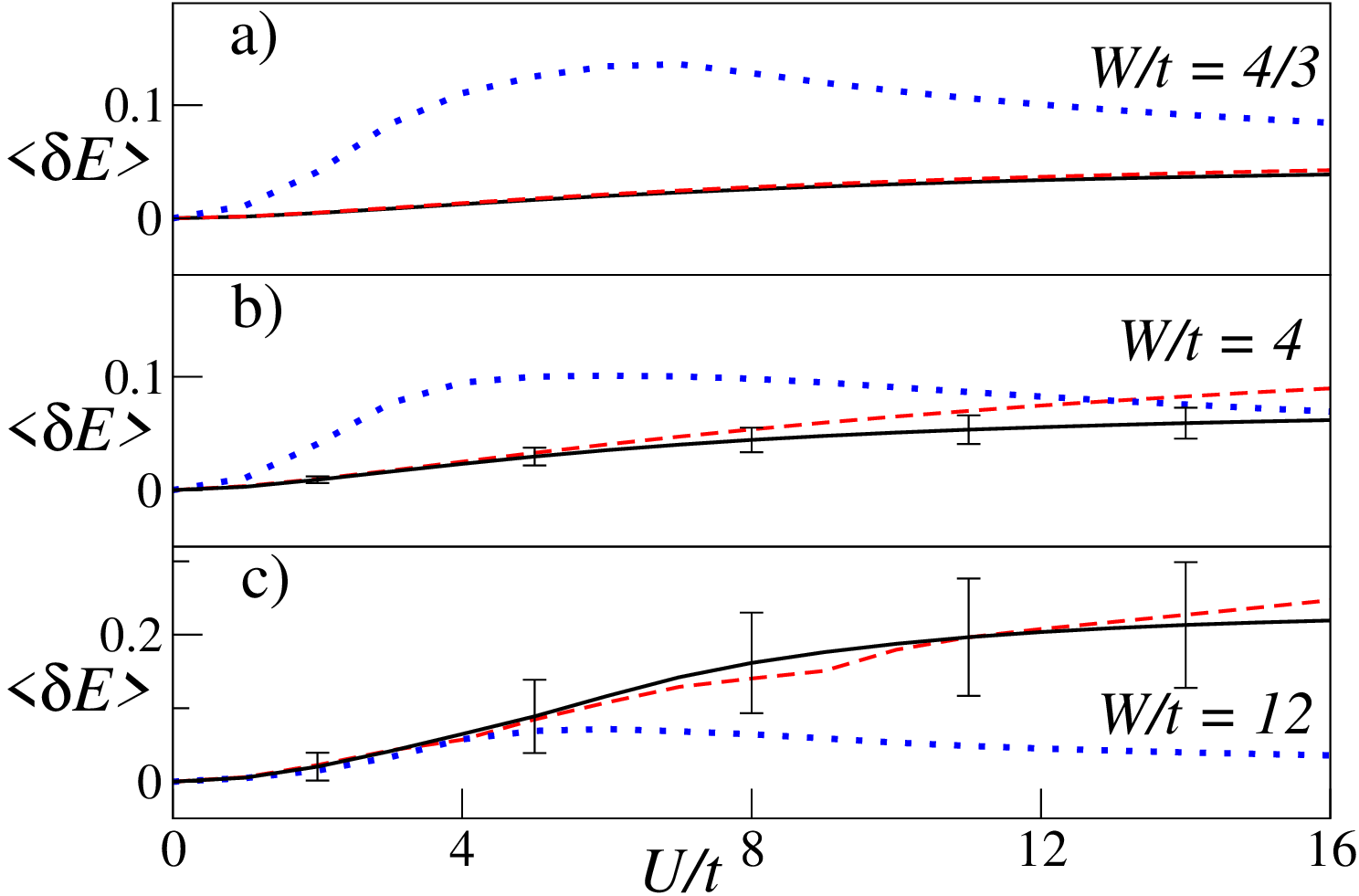}
\caption{\label{fig:10siteDIFnrgs6Ne10rand}(Colour online)
As in Fig.~\protect\ref{fig:10siteDIFnrgs10rand}, but for $N_e=6$ electrons.}
\end{center}
\end{figure}
Since there is no Mott transition for $N_e=6$, we expect the GWFs to be
valid over a larger range of $U$ than at half-filling.  This is supported by
Fig.~\ref{fig:10siteDIFnrgs6Ne10rand}, where the GWF energies are generally lower
than the UHF energies for weak and intermediate disorder.  As before, however, the GWF approximations are worse at large disorder.  It is interesting to note that $\langle \delta E\rangle_\mathrm{GWF}$ increases roughly linearly with $U$ at large $U$.  Similar behavior was found previously in the ordered case,\cite{yokoyama90}  and is consistent with the fact that $g_\mathrm{min}$ is larger for $N_e=6$ than it is for $N_e=10$ (i.e.\ that the projection of doubly-occupied states is less at $N_e=6$ than $N_e=10$).

\begin{figure}[tbp]
\begin{center}
%\vspace{0.5cm}
\includegraphics[clip=true,width=\columnwidth]{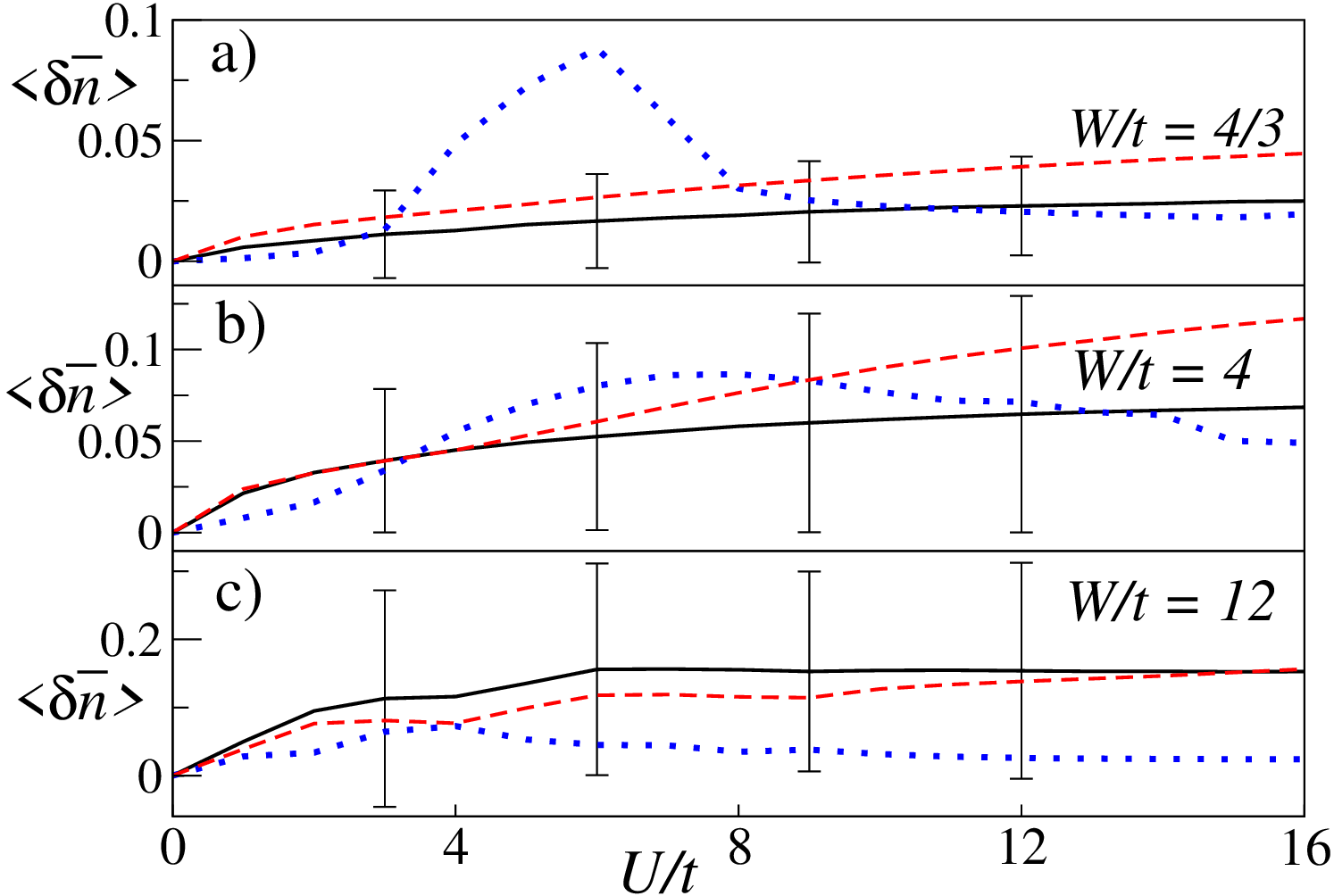}
\caption{\label{fig:10NiNe6_10rand}(Colour online)
As in Fig.~\protect\ref{fig:10Ni10rand}, but for $N_e=6$ electrons. Error bars are larger here than for $N_e=10$ because of the larger site-to-site variations in the accuracy of the GWF.}
\end{center}
\end{figure}

\begin{figure}[tbp]
\begin{center}
%\vspace{.4cm}
\includegraphics[clip=true,width=\columnwidth]{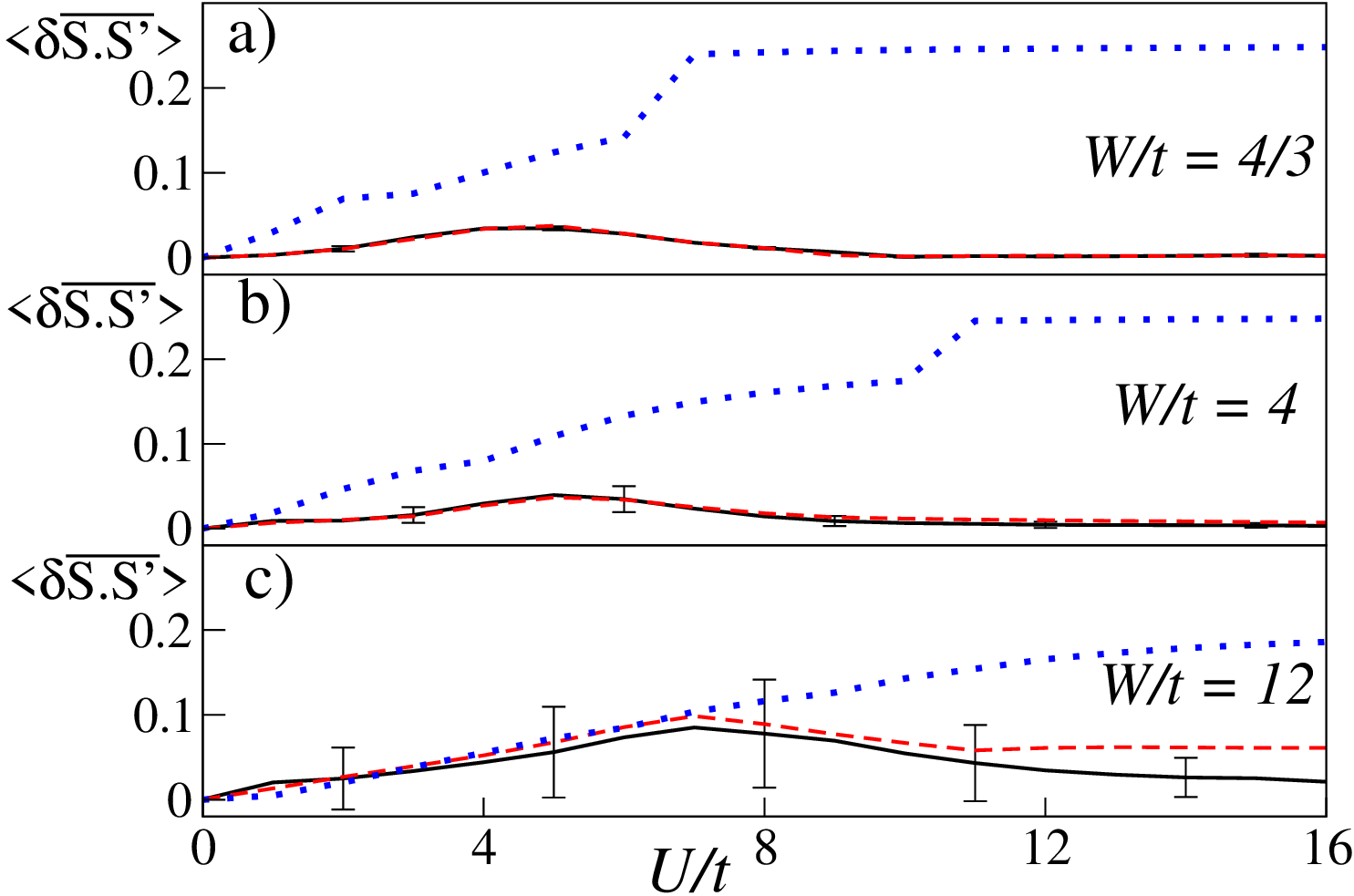}
\caption{\label{fig:10sisjNe6_10rand}(Colour online)
As in Fig.~\protect\ref{fig:10sisj10rand}, but for $N_e=6$ electrons.}
\end{center}
\end{figure}

The charge densities for $N_e=6$ (Fig.~\ref{fig:10NiNe6_10rand}) are similar to the half-filled case in the sense that the DFSGW charge densities are comparable to UHF\cite{chen08} at weak and intermediate disorder, but are less accurate than UHF at strong disorder. Similarly to the $N_e=10$ case, the {\it{average}} spin correlations (Fig.~\ref{fig:10sisjNe6_10rand}) are reasonably well reproduced at weak and intermediate disorder.

\section{Discussion}
\label{sec:discussion}

The  GWFs used in this work make two key simplifying assumptions:  first, that the correlation physics is approximately local and can be represented by a Gutzwiller partially-projected wavefunction and, second, that the spatial inhomogeneity associated with the disorder potential can be incorporated entirely within the uncorrelated product state.  In this section, we discuss these two assumptions, keeping in mind that, while it is possible to treat a large number of variational parameters using statistical sampling methods,\cite{pezzoli08} many of the interesting questions relating to disordered systems
require the ability to study large system sizes, and that it is therefore desirable to restrict the number of variational parameters as much as possible.  The results shown in this paper give some clues as to how this can be done.

There are two distinct regimes in our calculations, distinguished by the strength of the {\it{screened}} disorder potential, which in the DFSGW is  $W^\prime = W/\varepsilon_{min}$. For $W\lesssim D$, $W^\prime$ is sufficiently small that the GWFs used in this work give similar results for $\langle \delta E\rangle$  as
existing calculations for the ordered case (i.e. Fig. 4 of ~Ref.~\onlinecite{yokoyama90}).%as  existing calculations for the ordered case.
~In Ref.~\onlinecite{yokoyama90}, it was shown that the simplest GWF could be improved by inclusion of a long-range Jastrow factor, and the similarity of our results to theirs has a similar implication.  Interestingly, there is nothing in our results to suggest that adopting spatially-inhomogeneous variational parameters would make a significant improvement.  (Recall, Figs.~\ref{fig:comparisons_correlations_DFSGW_UHF_W4over3} and \ref{fig:comparisons_correlations_DFSGW_UHF_W4}, the accuracy with which the DFSGW approximation reproduces
the {\it{local}} charge density and spin correlations.)  It appears as if, for $W\lesssim D$, spatial inhomogeneity can be adequately incorporated in the product state provided that screening is included, either through a Hartree-Fock determination of the product state or through a variational parameter.  This represents a huge savings of computational effort since, if one makes a physically reasonable ansatz for the form of the Jastrow factor, the number of variational parameters can be reduced to four:  a screening parameter $\varepsilon$, a projection parameter $g$, and a pair of parameters for the Jastrow factor.\cite{yokoyama90,pezzoli08}

 For strong disorder, the situation
is different since spatial fluctuations %For strong disorder, the situation is different:  the elastic mean-free-path is a relevant length scale and spatial fluctuations
of the charge density are significant.  It appears that, in this case, spatial inhomogeneity cannot be included solely in $|\Psi_\mathrm{ps}\rangle$, but must be included in the variational parameters as well. %(Recall, for example, that for $W/t=12$, the GWF does a much better job of reproducing spatially-averaged properties than it does of reproducing the properties of a single site.)
This was the approach taken in Ref.~\onlinecite{pezzoli08} where the local variational parameters $g$ and $\varepsilon$ were allowed to vary from site to site  (but the nonlocal Jastrow factors were treated as spatially homogeneous in order to reduce the computational workload) on lattices of $N\sim 100$ sites.   It is unclear that any simplification of this approach is possible in the large-disorder limit.

\begin{figure}[tbp]
\begin{center}
%\vspace{0.2cm}
\includegraphics[clip=true,width=\columnwidth]{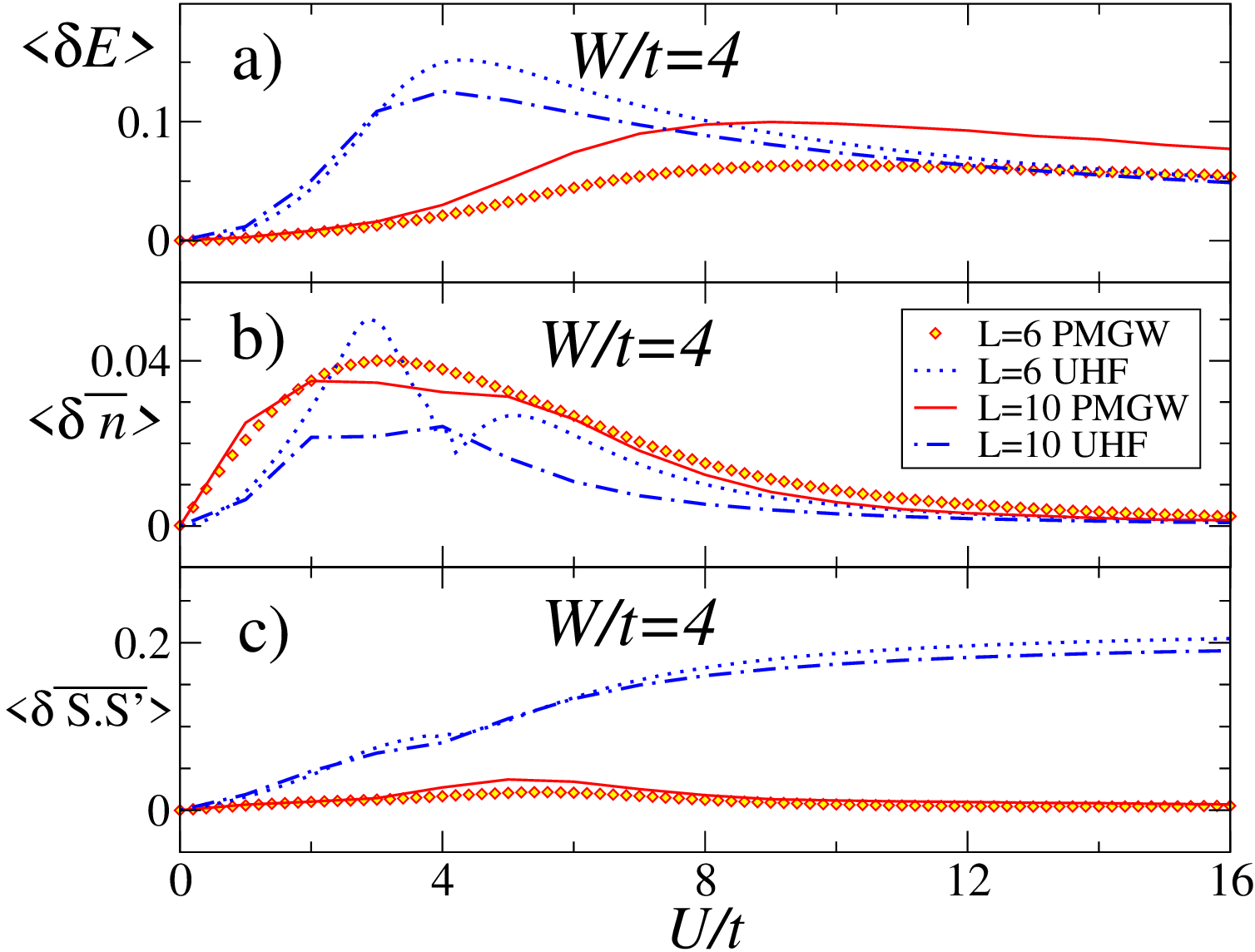}
\caption{\label{fig:10_6_comparison}(Colour online)
A comparison of the mean absolute difference between the
exact and the variational energies (a), local charge densities (b), and near-neighbour spin correlations (c).
The results are shown for ten-site clusters averaged over ten complexions of disorder (red solid lines for PMGW and blue dot-dashed lines for UHF), and six-site clusters averaged over fifteen configurations of disorder (diamonds for PMGW and blue dotted lines for UHF). Data are shown for step size $\Delta U = t$ for ten-site clusters and $\Delta U = 0.1t$ for six-site clusters.}
\end{center}
\end{figure}
Finally, we comment on finite size effects. We compare $\langle\delta E\rangle$, $\langle \delta {\overline n}\rangle$, and $\langle \delta {\overline{{\bf S}\cdot{\bf S'}}}\rangle$, for six and ten site clusters at half filling and intermediate disorder in Fig.~\ref{fig:10_6_comparison}. The GWF energy differences increase with system size and, in the weak disorder case, appear to scale towards the clean limit results shown in Fig.~4 of~Ref.~\onlinecite{yokoyama90} for large systems.We note that, while there are quantitative differences , $\langle \delta E\rangle$ for the $10$-site clusters has the same qualitative U-dependence as for the larger clean limit systems. More importantly for this work, $\langle \delta {\overline n}\rangle$ and $\langle \delta {\overline{{\bf S}\cdot{\bf S'}}}\rangle$ are nearly independent of system size. This is consistent with earlier scaling results in the disorder-free case\cite{kaplan82} and suggests that the high accuracy of the results in Fig.~\ref{fig:10_6_comparison} $(b)$ and $(c)$ will extend to larger systems. Thus, ability to benchmark GWF trial functions is not inhibited by finite size effects.

\section{Conclusions}
\label{sec:conclusions}
We have studied the Anderson-Hubbard model, one of the
simplest model Hamiltonians that can describe correlated
electrons moving on a disordered lattice.
We have compared the exact and variational ground states for disordered one dimensional
chains up to a length of 10 sites.   The main focus of our calculations was on finding the simplest variational states that accurately include the effects of disorder.  The quality of the approximations was determined by comparing a number of physical
quantities---the ground state energy, local charge densities, and near-neighbour spin correlations---to the exact results.  Because it has been, until recently, the standard tool for studying disordered systems, we also compared our variational states to the unrestricted Hartree-Fock approximation.

Our main conclusion is that for weak and intermediate disorder, an accurate description of the ground state can be obtained by taking a spatially inhomogeneous product state, and spatially homogeneous variational parameters.  This represents a significant reduction of computational workload over the more general case where both the product state and variational parameters must be inhomogeneous.

\appendix
\section{Two-Site Cluster}
\label{sec:Two-Site}
We compare the different product states for a two-site system, where an analytical solution can be found. For this cluster we choose the on-site energy of sites one and two to be
$V$ and $0$ respectively, and take $N_e=2$ (corresponding to half-filling). Then
\begin{eqnarray}\
\label{psiHF2sites}
|\psi_\mathrm{ps}\rangle&=&\frac{1}{\sqrt{N_{\uparrow}N_{\downarrow}}}(c_{1\uparrow}^{\dagger}+a_{\uparrow}c_{2\uparrow}^{\dagger})(c_{1\downarrow}^{\dagger}+a_{\downarrow}c_{2\downarrow}^{\dagger})|0\rangle\nonumber\\
&=&\frac{1}{\sqrt{1+a_\uparrow^2+a_\downarrow^2+(a_\uparrow a_\downarrow)^2}}\\
&&\times(|\uparrow\downarrow\rangle|0\rangle-a_\uparrow|\downarrow\rangle|\uparrow\rangle+a_\downarrow|\uparrow\rangle|\downarrow\rangle+a_\uparrow a_\downarrow|0\rangle|\uparrow\downarrow\rangle)\nonumber,
\end{eqnarray}
where $ {N_\sigma}^{-1/2}
\begin{pmatrix}
1\cr a_{\sigma}
\end{pmatrix}$ are normalized eigenvectors of the matrix
form of ${\cal H}_\mathrm{bl}$ for spin $\sigma$,
\begin{equation}
H_{\mathrm{bl},\sigma} = \begin{pmatrix}
V'+U{\bar{n}}_{1-\sigma} & -t            \cr
-t              & U{\bar{n}}_{2-\sigma} \cr
\end{pmatrix},
\end{equation}
with $V'=V/\varepsilon$,
\begin{eqnarray}
\label{defA}
a_{\sigma}&=&\frac{1}{2t}\{V'+U({\bar{n}}_{1-\sigma}-{\bar{n}}_{2-\sigma})\nonumber\\
&&+\sqrt{(U({\bar{n}}_{1-\sigma}-{\bar{n}}_{2-\sigma})+V')^2+4t^2}\},
\end{eqnarray}
and the many-particle states are defined as $|\uparrow\downarrow\rangle|0\rangle=c_{1\uparrow}^\dagger c_{1\downarrow}^\dagger|0\rangle$, $|\uparrow\rangle|\downarrow\rangle=c_{1\uparrow}^\dagger c_{2\downarrow}^\dagger|0\rangle$, etc.

Using Eqs.~(\ref{psiG-GW_0}) and (\ref{psiHF2sites}), $|\Psi_{{GWF}}\rangle$ is written as

\begin{widetext}
\begin{eqnarray}
&&|\Psi_{{GWF}}\rangle= \frac{1}{\sqrt{{1+a_{\uparrow}^2+a_{\downarrow}^2+(a_{\uparrow}a_{\downarrow})^2}}}
%\\&&\times
\left  ({g|\uparrow\downarrow\rangle|0\rangle-a_\uparrow|\downarrow\rangle|\uparrow\rangle+a_\downarrow|\uparrow\rangle|\downarrow\rangle+ga_\uparrow a_\downarrow|0\rangle|\uparrow\downarrow\rangle}
\right ),
%\nonumber
\label{GWvector}
\end{eqnarray}
and
\begin{eqnarray}
E_{{GWF}}=
\frac{g^2[(1+a_\uparrow^2a_\downarrow^2)U+2V]-2tg(a_\uparrow+a_\downarrow)(a_\uparrow a_\downarrow+1)+V(a_\uparrow^2+a_\downarrow^2)}{{g^2(1+(a_\uparrow a_\downarrow)^2)+a_\uparrow^2+a_\downarrow^2}}.
\end{eqnarray}
\end{widetext}

\begin{figure}[tbp]
\begin{center}
\vspace{0.7cm}
\includegraphics[width=\columnwidth]{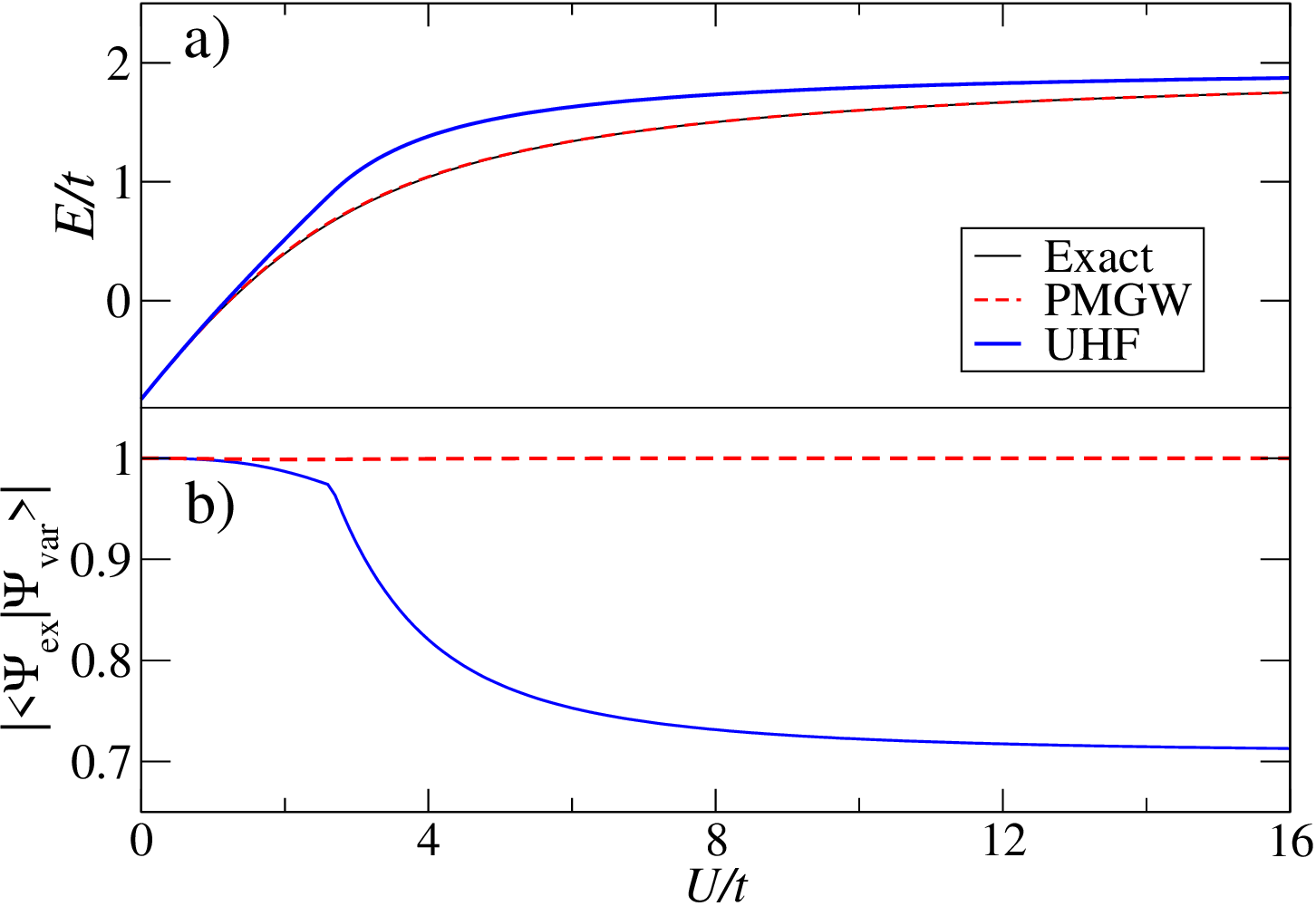}
\caption{\label{fig:2siteordered}(Colour online)
Variational solutions for the two-site Anderson-Hubbard model with $V=2.0t$.
(a) a comparison of the PMGW variational energy with
the exact and UHF ground-state energies.
The exact energies coincide with  the PMGW energies.
(b) the wavefunction overlap
$|\langle \Psi_{\textrm{ex}} | \Psi_{{\textrm{var}}} \rangle|$.}
\end{center}
\end{figure}

Numerical results for this cluster show that the DFSGW, PMGW$(g,\varepsilon)$, and PMGW energies are the same, and we therefore only show the results for PMGW in Fig. \ref{fig:2siteordered}. In this figure, we compare the magnitude of the PMGW energies with the UHF and exact energies. We also compare the overlaps of the PMGW and UHF wavefunctions with the exact wavefunctions. The overlaps are defined as $|\langle \Psi_{\textrm{ex}} | \Psi_{\textrm{\tiny{var}}} \rangle|$, where $\langle \Psi_{\textrm{ex}}|$ is the complex conjugate of the exact ground state eigenvector of $\mathcal{H}$.
The reason to study both ground state energies and overlaps is that a lower variational energy doesn't necessarily imply a better variational wave function.

In general, UHF energies are lower than PMHF energies (not shown), suggesting that the UHF approximation is better than the PMHF one. However, after the projection the situation is reversed. In Fig. \ref{fig:2siteordered} one sees that PMGW energies are exact, and that the UHF energies are higher than the PMGW energies for $U/t>2$, where magnetic moments develop in the UHF states.
Similarly, $|\langle \Psi_{ex}|\Psi_{PMGW}\rangle|=1$ while $|\langle \Psi_{ex}|\Psi_\mathrm{UHF}\rangle|$ decreases rapidly as moments develop.
This is consistent with the ordered case,\cite{yokoyama90} where the GWF is exact for the two-site problem, but is approximate for larger system sizes.

\acknowledgments
This work was supported in part by the NSERC of Canada.
\newpage
%\bibliographystyle{apsrev}
%\bibliography{myfileQ}% Produces the bibliography via BibTeX.

\begin{thebibliography}{52}
\bibitem[{Not()}]{NoteUHF}
\bibinfo{note}{Note that the most general decomposition includes spin-flip
  fields due to tilted moments. These fields have been shown to be important
  for some spin-glass phases, but are not relevant to the paramagnetic
  approximations used here. See Ref. \protect\onlinecite{chen08} for a complete
  discussion.}
\end{thebibliography}

\end{document}